\documentclass[journal]{IEEEtran}

\IEEEoverridecommandlockouts

\usepackage{amssymb}
\usepackage[cmex10]{amsmath}
\usepackage{stfloats}
\usepackage{graphicx}
\usepackage{subfigure}
\usepackage{tabularx}
\usepackage{epsfig,epsf,color,balance,cite}
\usepackage{verbatim}
\usepackage{url}
\usepackage{bm}
\usepackage{caption}
\usepackage{amsmath}
\usepackage{cases}
\usepackage{graphicx,xcolor,float}

\usepackage{bbm}

\newtheorem{lemma}{Lemma}

\usepackage{algorithm}
\usepackage{algorithmic}
\newtheorem{myDef}{Definition}

\usepackage{color}
\definecolor{myc1}{rgb}{0,0,0}
\definecolor{myc2}{rgb}{0,0,0}

\begin{document}
\title{Joint User Association and Beamforming in Multicast MmWave MIMO Networks: An RL-Based Approach}

\author{{Songling Zhang,} \emph{Member, IEEE},   {Nuocheng Yang,} \emph{Student Member, IEEE}, {Zhaohui Yang,} \emph{Member, IEEE},\\ {Danpu Liu,} \emph{Senior Member, IEEE}, {Zhilong Zhang,} \emph{Member, IEEE}, and  {Zhaoyang Zhang,} \emph{Fellow, IEEE}\\ 
\thanks{S. Zhang, Z. Yang, and Z. Zhang are with the College of Information Science and Electronic Engineering, Zhejiang University, and  Zhejiang Provincial Key Laboratory of Info. Proc., Commun. \& Netw. (IPCAN), Hangzhou, 310027, China, and also with Zhejiang Key Laboratory of Multimodal Communication Networks and Intelligent Information Processing (e-mails: \protect\url{sl-zhang@zju.edu.cn}; \protect\url{yang_zhaohui@zju.edu.cn}; \protect\url{ning_ming@zju.edu.cn} ).}
	\thanks{N. Yang, D. Liu, and Z. Zhang are with the Beijing Key Laboratory of Network System Architecture and Convergence, Beijing University of Posts and Telecommunications, Beijing, 100876, China (e-mail:  \protect\url{yangnuocheng@bupt.edu.cn}; \protect\url{dpliu@bupt.edu.cn};
 \protect\url{zhangzhilong@bupt.edu.cn}) .}
\thanks{\copyright~2026 IEEE. Personal use of this
	material is permitted. Permission from IEEE must
	be obtained for all other uses, in any current or
	future media, including reprinting/republishing
	this material for advertising or promotional
	purposes, creating new collective works, for
	resale or redistribution to servers or lists,
	or reuse of any copyrighted component of this
	work in other works.}
}

\maketitle

\begin{abstract}
In this paper, the problem of maximizing the sum rate of  multicast groups in a multi-base station (BS) cooperative millimeter wave (mmWave) multicast communication system is studied.  In the considered model, due to the real-time mobility of users, the users being served by a given BS and the hybrid beamforming matrices of BSs and users  are dynamic. Multiple BSs must cooperate to serve dynamic requests of multiple mobile users.  This problem is posed as an optimization framework whose goal is to maximize the sum rate of all  multicast groups by jointly optimizing the number of users served by the BSs and the hybrid beamforming matrices of both BSs and users. 
Since the hybrid beamforming matrices of both BSs and users
can be obtained for given BS-user associations and channel
conditions using the block diagonalization (BD) and fast optimization (FO) algorithms, the original
sum-rate maximization problem can be reduced to a
BS-user association optimization problem under the
adopted beamforming design.
To solve the resulting problem, we develop a value
decomposition based reinforcement learning (VD-RL) algorithm.
 Simulation results show that the proposed algorithm achieves up to a 50\% improvement in the sum rate of all multicast groups compared with the IDRL-BDFO scheme at a transmit power of 50 dBm.


\end{abstract}

\section{Introduction}

Driven by the vast available bandwidth, millimeter-wave (mmWave) technology has become a key enabling technology for modern wireless networks, enabling ultra-high data rates \cite{11540233,11535291,10138409, 6979963}. Nonetheless, these elevated frequency bands suffer from  much more severe path loss compared to their conventional lower-frequency counterparts \cite{11481588,8107556}. To compensate for this propagation bottleneck, massive multiple-input multiple-output (MIMO) techniques are typically leveraged to generate highly directional beams that guarantee robust link capacity \cite{10367817}. While fully digital architectures maximize the beamforming gain, equipping each antenna element with an individual radio frequency (RF) chain incurs unaffordable hardware costs and extreme power consumption. As a pragmatic paradigm, hybrid beamforming has gained substantial traction in mmWave systems, striking an appealing balance between minimizing RF chain requirements and preserving reliable signal-to-noise ratios (SNRs). Nevertheless, optimizing the aforementioned hybrid structures remains inherently challenging, largely due to the rigid constant-modulus restrictions imposed on the analog architectures and the tight coupling between the digital and RF precoding stages.

Extensive works have been dedicated to configuring hybrid beamforming  architectures in mmWave networks \cite{11481588,9110851,9933939,9110863,9369843,9882377,9646498,9910182,10192330,9454368,9743307,9894280,MmWaveDRLBeamforming2022,MultiRISMobileBeamTracking2026}. Fundamental multi-user hybrid architectures comprising analog and digital beamforming were investigated in \cite{9110851} and \cite{9933939}. To further enhance the multi-user designs, in \cite{11481588}, the authors proposed an innovative architecture that integrates rate-splitting multiple access  with hybrid beamforming for uplink mmWave communications. Meanwhile, \cite{9110863} extended these concepts to full-duplex multiple-relay configurations via a two-timescale approach. To address the practical challenges of channel acquisition and beam alignment, various solutions such as gridless estimation \cite{9369843}, efficient tracking mechanisms \cite{9882377}, and covariance-based designs accounting for beam squint \cite{9646498} have been developed for broadband and orthogonal frequency division multiplexing (OFDM)-based mmWave communication systems. Moreover, the integration of hybrid beamforming into specialized network paradigms has drawn significant attention, encompassing cell-free massive MIMO deployments \cite{9910182}, secure transmission for extended reality under  mixed digital-to-analog constraints \cite{10192330}, and simultaneous wireless information and power
transmission systems \cite{9454368}. Furthermore, recent advancements have explored the synergy between hybrid transceivers and reconfigurable intelligent surfaces (RIS), aiming to jointly optimize reflection matrices alongside analog-digital precoders to boost system performance and minimize the sum-mean-square-error \cite{9743307,9894280}.  
In addition, learning-based approaches have been
investigated to enhance hybrid beamforming in
dynamic mmWave environments. For example,
\cite{MmWaveDRLBeamforming2022} employed deep
reinforcement learning to design hybrid beamforming
matrices for high-mobility mmWave communications.
Furthermore, \cite{MultiRISMobileBeamTracking2026}
developed multi-agent reinforcement learning-based
beam tracking methods for multi-RIS-assisted
hybrid beamforming systems, enabling adaptive
beam management under user mobility. However, the majority
of prior works \cite{11481588,9110851,9933939,9110863,9369843,9882377,9646498,9910182,10192330,9454368,9743307,9894280,MmWaveDRLBeamforming2022,MultiRISMobileBeamTracking2026} primarily focus on hybrid
beamforming for unicast communication networks, where each base station (BS) sends unique data streams to different users. In contrast to
unicast, multicast enables the same content to be transmitted
to all users or a specific group of users using the same band
and modulation and coding scheme, which can improve the
spectral efficiency of wireless communication systems.


A substantial number of studies have explored beamforming techniques applied to mmWave multicast networks \cite{11174991,8546778,9014050,9685146,8417939,9145622,9767793, 9141433,8686225,8680660,9068231}. Among these, hybrid digital-to-analog beamforming designs have received prominent attention. For instance, fundamental hybrid architectures for multicast systems were investigated in \cite{11174991,9685146}, which were further extended to robust multi-group transmission strategies \cite{9767793} and joint transmit-receive precoder optimization frameworks \cite{9014050}. From the perspective of purely analog or low-complexity architectures, energy-efficient designs for single-group \cite{8686225} and generalized multigroup scenarios \cite{9068231}, alongside simplified transceiver algorithms \cite{9141433, 9145622}, have been widely studied. Furthermore, to enhance spectral efficiency, unified frameworks capable of simultaneously accommodating both unicast and multicast data streams have been proposed in \cite{8417939} and \cite{8680660}. From a network scheduling standpoint, the work in \cite{8546778} formulated an efficient multicast strategy leveraging multi-level codebooks, concurrent transmissions, and device-to-device (D2D) assistance.
However, these prior works \cite{11174991,8546778,9014050,9685146,8417939,9145622,9767793, 9141433,8686225,8680660,9068231} are confined to static user scenarios and cannot be readily extended to practical mobile environments. The integration of user mobility into mmWave multicast systems inherently complicates system design, introducing critical challenges such as dynamic BS-user  connections, adaptive user clustering, and the strict real-time constraints of hybrid beamforming optimization.


To address the challenges of dynamic user association,
various reinforcement learning-based approaches have been proposed.
In \cite{DynamicMmWaveMARL2020}, the authors developed
a multi-agent reinforcement learning framework
for adaptive user association in dynamic mmWave
networks based on local observations. A deep
Q-network-based user-centric association scheme
was proposed in \cite{UserCentricMmWaveDRL2021}
to accommodate time-varying ultra-dense mmWave
environments. Furthermore,
\cite{MmWaveAssociationHandoverRL2022} investigated
online reinforcement learning for load-balanced
user association and handover in mmWave-enabled
networks, while \cite{MobileNetworksMultiAgentQL2024}
extended multi-agent Q-learning to real-time user
association and handover in mobile networks by
accounting for coupled load-balancing constraints
and handover costs. These studies demonstrate
the potential of reinforcement learning for
adapting association policies to network dynamics.
Nevertheless, the aforementioned mobility-aware
approaches primarily address user association
or beam management in conventional cellular
transmission settings, rather than the joint
design of BS-user association and hybrid
transmit-receive beamforming for mobile
multigroup multicast networks. In the latter
setting, the achievable rate of each multicast
group depends on its weakest served user,
while the association decisions and beamforming
matrices are coupled through inter-group
interference and hardware constraints.
Therefore, efficiently coordinating dynamic
association and hybrid beamforming across
multiple BSs remains a challenging problem,
which motivates the proposed VD-RL-based
framework.

The main contribution of this paper is to develop a novel  mmWave multicast system that enables multiple BSs to serve users according to their mobility patterns, channel conditions, and request contents. The key contributions are summarized as follows:

\begin{itemize}
	
\item	
We consider a novel mmWave multicast system that considers mobile users. 
In this model, we tackle the sum-rate maximization problem for multicasting groups by jointly optimizing the BS-user associations, alongside both the transmit and receive beamforming matrices. The formulated optimization framework explicitly accounts for key practical constraints, including the maximum BS transmit power limitations, the user connection conditions, and the constant-amplitude constraints inherently required by the RF beamforming matrices.

\item
Since the hybrid beamforming matrices of both BSs and users
can be obtained for given BS-user associations and channel
conditions using the block diagonalization (BD) and fast
optimization (FO) algorithms, the original sum-rate
maximization problem can be reduced to a BS-user association
optimization problem under the adopted beamforming design.
To solve the resulting problem, we propose a novel value
decomposition-based reinforcement learning (VD-RL) algorithm.
Specifically, at each time slot, the BS-user associations
are first determined by the proposed VD-RL algorithm
based on the local observations and learned Q-value
functions of the BSs. Given the resulting associations
and channel conditions, the fully digital beamforming
matrices are then obtained using the BD method,
followed by the FO algorithm to design the corresponding
hybrid transmit and receive beamforming matrices.

\end{itemize}
Extensive numerical evaluations demonstrate the superiority of the proposed framework over conventional multi-agent baselines.
Specifically, at a transmit power of 50 dBm,
our scheme improves the sum rate of all multicast groups
by 50\% compared with the IDRL-BDFO scheme.

The paper proceeds as follows. Section II introduces the system model and optimization problem. The proposed BS-user association strategy and hybrid beamforming design are presented in Section III. Performance evaluations based on numerical simulations are shown in Section IV. Section V concludes the paper.

\begin{table*}[t]
	\renewcommand{\arraystretch}{1.5}
	\begin{center}
		{\color{myc2}{    \caption{List of Main Notation.}}}
		\label{tab:table1}
		\resizebox{\textwidth}{!}{
		\begin{tabular}{|c||c|c||c|}   \hline
			\textbf{Notation} & \textbf{Description} & \textbf{Notation} & \textbf{Description}\\
			\hline
			$ L$  & Number of  BSs &$K$ & Number of mobile users     \\   \hline
			$O$ & Number of contents &  $o_k $ & Content request of user $k$     \\   \hline
			$N^\textrm{U}$ & Number of antennas of each user&  $S^\textrm{U}$& Number of RF chains of each user   \\   \hline
			$\zeta$ & Data streams received by each user&  $p_{k}^\textrm{C}$ & Probability of user $k$ requesting  content $o_k$ at time $t$     \\   \hline
			$\boldsymbol{p}_{kt}$ & Possibility of the movement of each user at time $t$ &  $v_k$ &   Speed of user $k$  \\   \hline
			${\boldsymbol\phi _{lk}\left(t\right)} $ & Location of user $k$ of BS $l$ at time $t$ &  $\Delta t $ & Duration of each time slot $t$     \\   \hline
		$\boldsymbol{a}_l\left(t\right)$ & BS-user connection vector of BS $l$ at time $t$  &  $a_{lk}\left(t\right)  $ & Index of the connection between BS $l$ and user $k$   \\   \hline
			$\boldsymbol{a}\left(t\right)$ &     Actions of the BSs at time $t$ & $\mathcal U_{lo}\left(t\right)  $ & Set of users  that request content $o$ of BS $l$ at time $t$    \\   \hline
			$z_{lo}\left(t\right)$ & Index indicating whether users  of BS $l$  request content $o$ at time $t$ &   $\hat{\boldsymbol{s}}_{kl}\left(t\right)$ & Detected data of user $k$ of BS $l$ at time $t$   \\   \hline
			$\boldsymbol{s}_{e} $ & Data streams of content $e$ &   $\boldsymbol{F}_{b,e}^\textrm{B}$ & Baseband transmit beamforming matrix of group $e$ under BS $b$    \\   \hline
			$\boldsymbol{F}_{b}^\textrm{R}$ & RF
transmit beamforming matrix of BS $b$ & $\boldsymbol{H}_{l,k}\left(\boldsymbol{\phi}_{lk}\left(t\right)\right) $ & Effective channel between BS $l$ and user $k$    \\   \hline
			$\boldsymbol{W}_{kl}^{\textrm{R}}$ &  RF receive beamforming matrix of user $k$  of BS $l$  &   $\boldsymbol{W}_{kl}^{\textrm{B}}$  &  Baseband receive beamfoming matrix of user $k$ of BS $l$ \\   \hline
			$\boldsymbol{n}_{kl}\left(t\right)$ &  Additive white Gaussian noise vector of user $k$ & $\hat{s}_{kl,i}\left(t\right)$  &  Data stream $i$ of user $k$ from BS $l$ at the time $t$ \\   \hline
			${{\xi _{kl,i}\left(t\right)} }$ & SINR of data stream $i$ for the user $k$ at BS $l$  at  time $t$  & $I_{kl,i}\left(t\right)$ &  Interference from other streams  \\   \hline
			$J_{kl,i}\left(t\right)$ & Interference from other groups & $c_{kl}\left(t\right)$ & Achievable data rate of user $k$  of BS $l$ at time $t$    \\   \hline
		$W$ &  Bandwidth of each BS &  $c_{l{o}}\left(t\right)$  & Data rate for the group involving user $k$ at BS $l$   \\   \hline
			$D$ & Maximum transmit power of each BS  &  $\boldsymbol\xi _{t} $ & States of the agents at time $t$  \\   \hline
			$\boldsymbol\xi _{lt}  $ & State of BS $l$ at time $t$ &  $E_{l} $ & Maximum user index  of BS $l$    \\   \hline
			$\boldsymbol\xi _{lkt}$ & State of user $k$  of BS $l$ at time $t$ &  $\boldsymbol{B}_{l,o}\left(t\right)$ & Fully digital transmission beamforming matrix  \\   \hline
		$\boldsymbol{J}_{kl}\left(t\right)$ & User $k$'s fully digital reception beamforming matrix at BS $l$  &  $N_L $ & Maximum group under BS $L$     \\   \hline
			$\bar{\boldsymbol{H}}_{l{\left (o-1\right )}}$& Effective channels for all members in group $o$-1 of BS $l$ &  $E_{l{\left (o-1\right )}}$ &  Number of members of group $o$-1    \\   \hline
			${\boldsymbol{H}}_{lo,k}$ & Effective channels for user $k$ in group $o$ of BS $l$ & ${{\boldsymbol{G}}_{lo}}$ &  Power distribution matrix for group $o$ in BS $l$     \\   \hline
			$p_i$ & Transmit power of data stream $i$ &  $r\left( \boldsymbol\xi _{t}, \boldsymbol{a}\left(t\right)    \right) $& Total reward of the BSs   \\   \hline
			$r_{l}\left( \boldsymbol\xi _{lt}, \boldsymbol{a}_{l}\left(t\right)    \right)$ &Reward of BS $l$ at time $t$ &   $\rho   $ & Penalty for multiple BSs serving one user  \\   \hline
			$Q_{\boldsymbol{\vartheta}_{lt}} \left( \boldsymbol\xi _{lt}, \boldsymbol{a}_{l}\left(t\right) \right) $ & Individual $Q$ value function & $\boldsymbol{\vartheta}_{lt}$& Parameter of deep neural network     \\   \hline
		$Q_{tot} \left( \boldsymbol\xi _{t}, \boldsymbol{a}\left(t\right) \right)$ & Global $Q$ value function  &   $A\left( \boldsymbol{\vartheta}_{1t}, \ldots,  \boldsymbol{\vartheta}_{Lt} \right) $ &  loss function of the global $Q$ value function  \\   \hline
  $\lambda_{\boldsymbol{\vartheta}_{lt}}$ & Updating rate  &   $\nabla_{\boldsymbol{\vartheta}_{lt}} A\left( \boldsymbol{\vartheta}_{1t}, \ldots,  \boldsymbol{\vartheta}_{Lt} \right) $ &  Gradient of the global $Q$ value function \\   \hline
		\end{tabular}}
	\end{center}
\end{table*}

\section{System Model and Problem Formulation}\label{SYSTEM MODEL AND PROBLEM FORMULATIO}

\subsection{System Model}\label{S1}
\begin{figure}[t]
	\centering
	\includegraphics[width=9cm]{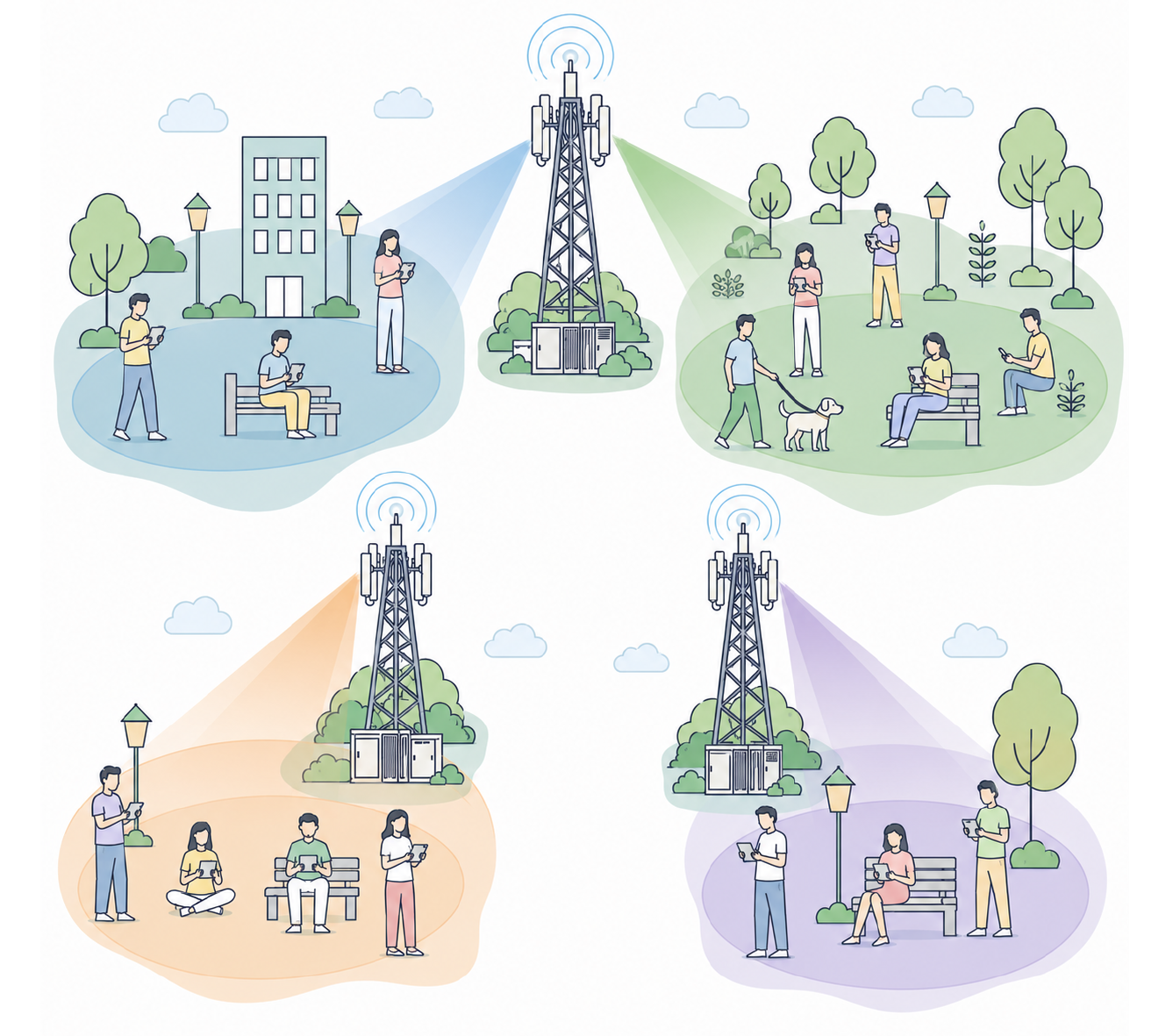}
	\caption{ The considered multigroup  multicast MIMO communication system.}\label{fig 1}
\end{figure}
As illustrated in Fig. \ref{fig 1}, we investigate a multigroup multicast MIMO system comprising $L$ BSs, indexed by the set $\mathcal{L}$, which jointly serve  $K$ mobile users. Each user follows a probabilistic mobility pattern and  requests  identical content in distinct time slots. Specifically, we postulate that the $k$-th user is interested in  content $o_k$  from a predefined content set $\mathcal{O}$ of  $O$ contents. At any given time slot $t$, the probability of user $k$ actively requesting content $o_k$ is denoted by $p_{k}^\textrm{C}$, with the complementary probability $1-p_{k}^\textrm{C}$ indicating an idle state for that specific content. It is worth noting that this  framework can be seamlessly generalized to scenarios involving time-varying user preferences. During each time slot $t$, all users are dynamically clustered into distinct multicast groups in accordance with their instantaneous spatial distributions and  content demands. Regarding the transceiver hardware, each BS utilizes an array of $N^\textrm{B}$ antennas coupled with $O\zeta$ RF chains to support downlink transmission. Furthermore, each user utilizes $N^\textrm{U}$ antennas and $S^\textrm{U}$ RF chains to  recover $\zeta$ spatial data streams. 
The main notations of this work are listed in
Table I.

\subsubsection{Mobility Model}
To characterize the spatio-temporal dynamics of the users, we adopt a discrete-time 2D random walk mobility model. In particular, during any given time interval, a user may either stay stationary or transition towards one of the four orthogonal directions. The probabilistic movement state of user $k$ at time $t$ is encapsulated by the vector $\boldsymbol{p}_{kt} = \left[p_{kt,0},p_{kt,1}, p_{kt,2},p_{kt,3},p_{kt,4}\right]$. Herein, $p_{kt,0}$ defines the probability of remaining at the current position, whereas the elements $p_{kt,1}$ through $p_{kt,4}$ correspond to the probabilities of moving upward, downward, leftward, and rightward, respectively. Assuming a uniform velocity $v_k$ and a constant time slot duration $\Delta t$, the instantaneous Cartesian coordinate ${\boldsymbol\phi _{lk}\left(t\right)} = [{\phi _{lkt,1}},{\phi _{lkt,2}}]$ of user $k$ within the service area of BS $l$ evolves according to the following stochastic spatial process:
\begin{equation}\label{System model1}
\boldsymbol\phi_{lk}\left(t+1\right)=\left\{\begin{array}{l}
{\left[\phi_{lk t, 1}, \phi_{lk t, 2}\right], \quad \text{w.p. } p_{k t+1,0},} \\
{\left[\phi_{lk t, 1}, \phi_{lk t, 2}+v_k \Delta t\right], \quad \text{w.p. } p_{k t+1,1},} \\
{\left[\phi_{lk t, 1}, \phi_{lk t, 2}-v_k \Delta t\right], \quad \text{w.p. } p_{k t+1,2},} \\
{\left[\phi_{lk t, 1}-v_k \Delta t, \phi_{lk t, 2}\right], \quad \text{w.p. } p_{k t+1,3},} \\
{\left[\phi_{lk t, 1}+v_k \Delta t, \phi_{lk t, 2}\right], \quad \text{w.p. } p_{k t+1,4}.}
\end{array}\right.
\end{equation}
The overall VD-RL-based framework can also be extended to other mobility models, such as random waypoint and Gauss–Markov models. Such extensions would involve adopting the corresponding mobility dynamics in the environment and training or fine-tuning the association policy accordingly. Where necessary, the state representation can be augmented with additional mobility information, such as velocity or direction, to account for temporal correlations.

\subsubsection{Channel Model}

Following the geometric multipath channel model adopted in our prior work \cite{10196368}, we characterize each BS--user channel as a superposition of line-of-sight (LOS) and non-line-of-sight (NLOS) propagation paths.
 Both the BSs and users employ uniform
linear arrays (ULAs). At time slot $t$, the channel
between BS $l$ and user $k$ is expressed as

\begin{equation}
\label{eq:channel_model}
\begin{split}
&\boldsymbol{H}_{l,k}
\left(\boldsymbol{\phi}_{lk}\left(t\right)\right)\\
& =
\sqrt{\frac{N^\textrm{B}N^\textrm{U}}{N_\textrm{p}}}
\sum_{p=1}^{N_\textrm{p}}
g_{lk,p}\left(t\right)
\times
\boldsymbol{a}_{N^\textrm{U}}
\left(\theta_{lk,p}\left(t\right)\right)
\boldsymbol{a}_{N^\textrm{B}}^{\textrm{H}}
\left(\varphi_{lk,p}\left(t\right)\right),
\end{split}
\end{equation}
where $N_\textrm{p}$ denotes the total number of
propagation paths, including both LOS and NLOS paths.
The coefficient $g_{lk,p}\left(t\right)$ denotes the
effective complex gain of path $p$, including
propagation attenuation and antenna-gain scaling.
The variables $\theta_{lk,p}\left(t\right)$ and
$\varphi_{lk,p}\left(t\right)$ denote the corresponding
angle of arrival and angle of departure, respectively.
For a ULA with $N$ antennas, the normalized array
response vector is given by

\begin{equation}
\label{eq:array_response}
\begin{split}
\boldsymbol{a}_{N}\left(\theta\right)
={}&\frac{1}{\sqrt{N}}
\left[
1,
e^{\textrm{j}\frac{2\pi d_\textrm{a}}{\lambda}\sin\theta},
\ldots,
e^{\textrm{j}\frac{2\pi d_\textrm{a}}{\lambda}
\left(N-1\right)\sin\theta}
\right]^{\textrm{T}},
\end{split}
\end{equation}
where $d_\textrm{a}$ is the antenna spacing,
$\lambda$ is the signal wavelength, and
$\textrm{j}=\sqrt{-1}$.  In this paper, we assume that perfect channel state information at the transmitter (CSIT) is available.

\subsubsection{Data Rate Model}
To mathematically formalize the network topology, we define the binary association vector for BS $l$ at time slot $t$ as $\boldsymbol{a}_l\left(t\right)=\left[a_{l1}\left(t\right),\ldots,a_{lK}\left(t\right)\right]$. Specifically, the Boolean indicator $a_{lk}\left(t\right) \in \left\{0,1\right\}$ takes the value of $1$ if user $k$ is currently served by BS $l$, and evaluates to $0$ otherwise. To ensure that no user is simultaneously associated with multiple BSs, we impose the at-most-one-BS association constraint
$\sum_{l=1}^{L} a_{lk}\left(t\right) \leq 1$ for each user $k$. In a similar vein, the instantaneous user preferences are  represented by the content demand vector $\boldsymbol{y}_o\left(t\right)=\left[y_{o1}\left(t\right),\ldots,y_{oK}\left(t\right)\right]^{\textrm{T}}$. The element $y_{ok}\left(t\right) \in \left\{0,1\right\}$ acts as a request indicator variable, taking the value of 1 when user $k$ requests content $o$ at time slot $t$, and 0 otherwise.

Given that each user requests one content from the set $\mathcal{O}$, BS $l$ can categorize its users into a maximum of $O$ groups according to their content requests during each time slot. Let $z_{lo}(t) \in \{0,1\}$ serve as a binary indicator, which is set to $1$ if at least one user linked to BS $l$ requests content $o$ at time slot $t$, and $0$ if not. The relationship between the user association vector $\boldsymbol{a}_l(t)$ and the content request indicator $z_{lo}(t)$ is represented as
\begin{equation}
z_{lo}\left(t\right) =  \mathcal X_{\boldsymbol{a}_l\left(t\right) \boldsymbol{y}_o\left(t\right) \neq 0},
\end{equation} 
where $\mathcal X$ denotes the characteristic function. For a specific $\boldsymbol{a}_l(t)$, the data recovered for user $k$ served by BS $l$ during time slot $t$ can be expressed as
\begin{equation}\label{System model1}
\begin{split}
&\hat{\boldsymbol{s}}_{kl}\left(t\right)\! 
 \\& =\sum_{b=1 }^{L}  \sum_{e=1 }^{O} z_{be}\left(t\right)\left(\boldsymbol{W}_{kl}^{\textrm{B}}\right)^{\textrm{H}} \left(\boldsymbol{W}_{kl}^{\textrm{R}}\right)^{\textrm{H}} \boldsymbol{H}_{b,k}\left(\boldsymbol{\phi}_{lk}\left(t\right)\right)
\boldsymbol{F}_{b,e}^\textrm{R} \boldsymbol{F}_{b,e}^\textrm{B} \boldsymbol{s}_{e} \\&
+\left(\boldsymbol{W}_{kl}^{\textrm{B}}\right)^{\textrm{H}} \left(\boldsymbol{W}_{kl}^{\textrm{R}}\right)^{\textrm{H}}\boldsymbol{n}_{kl}\left(t\right),
\end{split}
\end{equation}  
where $\boldsymbol{s}_{e}=\left[s_{e,1},\ldots, s_{e,\zeta}\right]^{\textrm{T}} \in \mathbb{C}^{\zeta \times 1}$ denotes the transmitted data streams intended for content $e$, which is assumed to be normalized  such that $\mathbb{E} \left[\boldsymbol{s}_{e} \boldsymbol{s}_{e}^{\textrm{H}}\right]=\boldsymbol{I}_{\zeta}$. For the hybrid beamforming architecture at the transmitter, the data streams serving group $e$ from BS $b$ are successively processed by the digital baseband precoder $\boldsymbol{F}_{b,e}^\textrm{B}$ and the analog RF precoder $\boldsymbol{F}_{b,e}^\textrm{R}$. The precoded signals then propagate through the spatial environment, characterized by the effective channel matrix $\boldsymbol{H}_{b,k}\left(\boldsymbol{\phi}_{lk}\left(t\right)\right)$ linking the $b$-th BS to the $k$-th user. At the receiver side, user $k$ situated in the coverage area of BS $l$ decodes the incoming signals utilizing a hybrid combining architecture, which comprises the analog RF combiner $\boldsymbol{W}_{kl}^{\textrm{R}}$ followed by the digital baseband combiner $\boldsymbol{W}_{kl}^{\textrm{B}}$. Furthermore, the additive white Gaussian noise  vector at user $k$ associated with BS $l$ is denoted by $\boldsymbol{n}_{kl}\left(t\right) \in \mathbb{C}^{N^{\textrm{U}} \times 1}$, whose entries are modeled as independent and identically distributed  complex Gaussian random variables. Consequently, the received signal formulated in (\ref{System model1}) can be  partitioned into the desired data component and the noise component. Therefore, at time slot $t$, the $i$-th demodulated data stream for user $k$ connected to BS $l$ is mathematically expressed as
\begin{equation}\label{System model2}
\begin{split}
&\hat{s}_{kl,i}\left(t\right) \\&= \left(\boldsymbol{w}_{kl,i}^{\textrm{B}}\right)^{\textrm{H}} \left(\boldsymbol{w}_{kl,i}^{\textrm{R}}\right)^{\textrm{H}} \boldsymbol{H}_{l,k}\left(\boldsymbol{\phi}_{lk}\left(t\right)\right) \boldsymbol{F}_{l,o_{k},i}^\textrm{R} \bar{\boldsymbol{f}}_{l,o_{k},i}^{\textrm{B}}  s_{o_{k},i} \\
&+ \sum_{j=1,j \neq i}^{\zeta} \left(\boldsymbol{w}_{kl,i}^{\textrm{B}}\right)^{\textrm{H}} \left(\boldsymbol{w}_{kl,i}^{\textrm{R}}\right)^{\textrm{H}} \boldsymbol{H}_{l,k}\left(\boldsymbol{\phi}_{lk}\left(t\right)\right) \boldsymbol{F}_{l,o_{k},j}^\textrm{R} \bar{\boldsymbol{f}}_{l,o_{k},j}^{\textrm{B}}  s_{o_{k},j} \\
&+ \sum_{b=1,b\neq l}^{L} \sum_{e=1,e\neq o_k}^{O}z_{be}\left(t\right)\left(\boldsymbol{W}_{kl}^{\textrm{B}}\right)^{\textrm{H}} \left(\boldsymbol{W}_{kl}^{\textrm{R}}\right)^{\textrm{H}} \boldsymbol{H}_{b,k}\left(\boldsymbol{\phi}_{lk}\left(t\right)\right)
\boldsymbol{F}_{b,e}^\textrm{R}  \\  & \ \  \ \, \boldsymbol{F}_{b,e}^\textrm{B} \boldsymbol{s}_{e}\\
&+ \left(\boldsymbol{w}_{kl,i}^{\textrm{B}}\right)^{\textrm{H}} \left(\boldsymbol{w}_{kl,i}^{\textrm{R}}\right)^{\textrm{H}} \boldsymbol{n}_{kl,i}\left(t\right),
\end{split}
\end{equation}
where $o_{k}$ indicates the specific content requested by user $k$, which satisfies the condition $y_{ok}\left(t\right)=1$. To facilitate   signal analysis, we define $\boldsymbol{w}_{kl,i}^{\textrm{B}}$ as the $i$-th row vector of the baseband combiner $\boldsymbol{W}_{kl}^{\textrm{B}}$, and $\bar{\boldsymbol{f}}_{l,o_k,i}^{\textrm{B}}$ as the $i$-th column vector of the baseband precoder $\boldsymbol{F}_{l,o_k}^\textrm{B}$. By dissecting the composition of (\ref{System model2}), the received signal is mathematically decomposed into four distinct components: the intended data stream, the inter-stream interference within the same group, the inter-group multi-user interference, and the background noise. Building upon this signal decomposition, the instantaneous signal-to-interference-plus-noise ratio (SINR) for user $k$ decoding the $i$-th data stream from BS $l$ at time slot $t$ is formulated as

\begin{equation}\label{R-}
\begin{split}
{{\xi _{kl,i}\left(t\right)} }&= \\
&\!\!\!\!\!\!\!\!\!\!\!\!\!\! \frac{ \left | \left(\boldsymbol{w}_{kl,i}^{\textrm{B}}\right)^{\textrm{H}} \left(\boldsymbol{w}_{kl,i}^{\textrm{R}}\right)^{\textrm{H}} \boldsymbol{H}_{l,k}\left(\boldsymbol{\phi}_{lk}\left(t\right)\right) \boldsymbol{F}_{l,o_{k},i}^\textrm{R} \bar{\boldsymbol{f}}_{l,o_{k},i}^{\textrm{B}}\right |^2}       {{I_{kl,i}\left(t\right)}+{J_{kl,i}\left(t\right)}+\sigma^2},
\end{split}
\end{equation}
 where $I_{kl,i}\left(t\right)= \sum_{j=1,j \neq i}^{\zeta}  \left | \left(\boldsymbol{w}_{kl,i}^{\textrm{B}}\right)^{\textrm{H}} \left(\boldsymbol{w}_{kl,i}^{\textrm{R}}\right)^{\textrm{H}} \boldsymbol{H}_{l,k}\left(\boldsymbol{\phi}_{lk}\left(t\right)\right) \notag \right.\\ \left.      \boldsymbol{F}_{l,o_{k},j}^\textrm{R}  \bar{\boldsymbol{f}}_{l,o_{k},j}^{\textrm{B}}\right |^2$ denotes the inter-stream interference caused by the other simultaneously transmitted spatial streams within the same multicast group, and $J_{kl,i}\left(t\right)= \sum_{b=1,b\neq l}^{L} \sum_{e=1,e\neq o_k}^{O} \left | z_{be}\left(t\right)\left(\boldsymbol{W}_{kl}^{\textrm{B}}\right)^{\textrm{H}} \left(\boldsymbol{W}_{kl}^{\textrm{R}}\right)^{\textrm{H}} \boldsymbol{H}_{b,k}\left(\boldsymbol{\phi}_{lk}\left(t\right)\right)   \right.\\ \left. 
\boldsymbol{F}_{b,e}^\textrm{R} \boldsymbol{F}_{b,e}^\textrm{B}\right |^2$ denotes the inter-group interference caused by transmissions intended for other multicast groups.
The instantaneous achievable data rate for the $k$-th user served by BS $l$ during time slot $t$ can be formulated as
\begin{equation}\label{R}
\begin{split}
c_{kl}\left(t\right) =   W\sum_{i=1}^{\zeta} \log_2 \left ( 1 +   
   \xi _{kl,i}\left(t\right)  \right), 
\end{split}
\end{equation}
where $W$ denotes  the bandwidth.

\addtolength{\topmargin}{0.02in}
 
Inherently, the multicast transmission paradigm dictates that the  data rate for a specific group served by BS $l$ is bottlenecked by the worst-case user within that group. Therefore, the effective group rate is strictly bounded by the minimum individual achievable rate, expressed mathematically as
\begin{equation}\label{System model}
\begin{split}
c_{l{o}}\left(t\right)= 
\mathop {\min }\limits_{k \in {\mathcal U_{l,o}\left(t\right)}} \left\{c_{kl}\left(t\right) \right\},
\end{split}
\end{equation}
where ${\mathcal U_{l,o}\left(t\right)}$ represents the set of users requesting content $o$ served by BS $l$ at time $t$.

\subsection{Problem Formulation}\label{S123}
Having defined the  achievable rate, we now formally formulate the optimization problem.  The formulated optimization problem aims to maximize  the system sum rate across all multicast groups through the joint optimization of the  hybrid transmit precoders $\boldsymbol{F}_{l}^\textrm{R}\left(t\right)$, $\boldsymbol{F}^\textrm{B}_{{l,o}}\left(t\right)$, the hybrid receive combiners $\boldsymbol{W}_{kl}^{\textrm{R}}\left(t\right)$, $\boldsymbol{W}_{kl}^{\textrm{B}}\left(t\right)$, and the dynamic user association indicators $a_{lk}\left(t\right)$. Consequently, the  sum rate maximization problem is mathematically cast as follows

\begin{subequations}   \label{PF1}
	\begin{align}
	 &\mathop {\max}_{\boldsymbol{F}_{l}^\textrm{R}\left(t\right), \boldsymbol{F}^\textrm{B}_{{l,o}}\left(t\right), \boldsymbol{W}_{kl}^{\textrm{R}}\left(t\right), \boldsymbol{W}_{kl}^{\textrm{B}}\left(t\right),a_{lk}\left(t\right) }     \mathbb E \left(\sum\limits_{l = 1}^L \sum\limits_{o = 1}^{O} z_{lo}\left(t\right) { c_{lo}\left(t\right) }\right) \tag{\theequation} \\
  	&  \textrm{s.t.}	 \ \   \sum_{l=1 }^{L} a_{lk}\left(t\right) \leq 1, \forall k,\\
	&  \  \; \ \quad \sum\limits_{o = 1}^{O} z_{lo}\left(t\right)\left\| \boldsymbol{F}_{l,o}^\textrm{R}\left(t\right)\boldsymbol{F}_{l,o}^\textrm{B}\left(t\right)\right\|_F^2 \le D,  \forall l,\\
    & \ \ \; \quad {\rm{      }}\left| \boldsymbol{F}_{l}^\textrm{R}\left(i,j\right)\left(t\right) \right| = \left| \boldsymbol{W}_{kl}^{\textrm{R}}\left(i,j\right)\left(t\right) \right| = 1, \forall i,j,
	\end{align}
\end{subequations}
where $D$ denotes the highest allowable transmission power for each BS.
Constraint (\ref{PF1}a) ensures that each user is associated with at most one BS
in each time slot. Constraint (\ref{PF1}b) bounds the transmit power of each BS by the maximum allowance $D$, while (\ref{PF1}c) imposes amplitude restrictions on the RF beamforming matrices at both the transceivers. Due to the tight coupling among the optimization variables in the objective function and the inherent non-convexity of the feasible region defined by (\ref{PF1}a)-(\ref{PF1}c), problem (\ref{PF1}) is highly non-convex and mathematically intractable in its current form.

 \section{Proposed  Scheme}\label{Proposed Value Decomposition-Reinforcement Learning Algorithm}



Since the hybrid beamforming matrices of both BSs and users can be
obtained for given BS-user associations and channel conditions
using the BD and FO algorithms, the original sum-rate maximization
problem can be reduced to a BS-user association optimization
problem under the adopted beamforming design. To solve the
resulting problem, we propose a VD-RL algorithm. Specifically,
the BS-user associations are determined by the proposed VD-RL
algorithm based on the local observations and learned Q-value
functions of the BSs. Given the resulting associations and
channel conditions, we employ a BD
method to derive the fully digital transmit and receive
beamforming matrices for the BSs and users, respectively.
Based on the obtained fully digital solutions, a low-complexity
FO algorithm is then employed to design
the hybrid transmit precoders
$\boldsymbol{F}_{l}^{\textrm{R}}(t)$,
$\boldsymbol{F}^{\textrm{B}}_{{l,o}}(t)$
and receive combiners
$\boldsymbol{W}_{kl}^{\textrm{R}}(t)$,
$\boldsymbol{W}_{kl}^{\textrm{B}}(t)$
at time slot $t$.
The main motivation for combining VD-RL with BD and FO is that the BS-user association and beamforming variables have different mathematical characteristics. Specifically, VD-RL is employed to dynamically determine the discrete BS-user associations according to the time-varying user locations and content requests, while enabling cooperative decision-making among BSs through value decomposition. Given the BS-user associations, the BD method exploits the channel structure to construct fully digital transmit and receive beamformers, while the FO algorithm subsequently approximates these matrices to obtain feasible hybrid beamformers.

\subsection{VD-RL Algorithm.}\label{Value Decomposition Based Reinforcement Learning
Algorithm}



\begin{figure}[t]
	\centering
	\includegraphics[width=9cm]{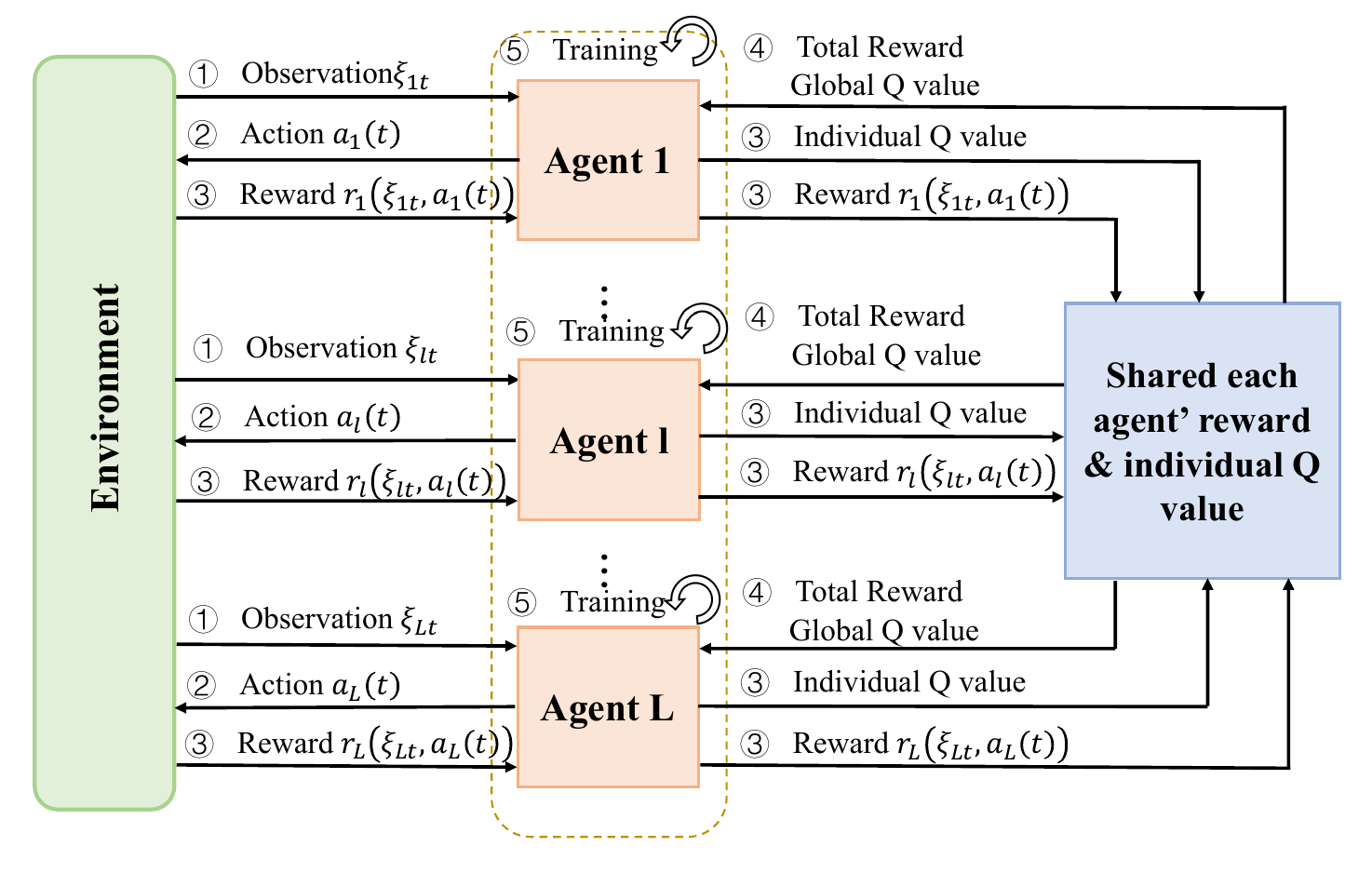}
	\caption{ The architecture of VD-RL.}\label{VDRL}
\end{figure}

The architecture of the proposed VD-RL framework is illustrated in Fig. \ref{VDRL}. First, we elaborate on its fundamental components. Subsequently, 
we demonstrate how the proposed VD-RL algorithm is applied to solve the BS-user association problem.

\subsubsection{VD-RL
Components}

The design of the VD-RL scheme encompasses six essential components.

\begin{itemize}
\item 
Agents:  In the proposed VD-RL framework, the BSs act as  agents to  determine the dynamic user association.

\item

States: The local state of BS $l$ at time slot $t$ is represented by the vector $\boldsymbol{\xi}_{lt}$, defined as $\boldsymbol{\xi}_{lt} = \left[\boldsymbol{\xi}_{l1t}, \ldots, \boldsymbol{\xi}_{lkt}, \ldots, \boldsymbol{\xi}_{lE_{l}t}\right]$. The parameter $E_{l}$ denotes the count of users under the coverage of BS $l$. For each user $k$, the state vector $\boldsymbol{\xi}_{lkt} = \left[{\boldsymbol{\phi}_{lk}(t)}, o_k\right]$ describes the user's state at time $t$ within the area of BS $l$. The element ${\boldsymbol{\phi}_{lk}(t)} = [{\phi_{lkt,1}}, {\phi_{lkt,2}}]$ indicates the user's position in two dimensions, while $o_k$ signifies the content request. During time slot $t$, BS $l$ can only access the state information $\boldsymbol{\xi}_{lt}$ for users within its coverage area.

\item

Actions: The action taken by each agent dictates the selection of active users to be served. At time slot $t$, the decision of BS $l$ is denoted by $\boldsymbol{a}_l(t) = \left[a_{l1}(t), \ldots, a_{lK}(t)\right]$, where the binary indicator $a_{lk}(t) \in \{0,1\}$ specifies whether user $k$ is associated with BS $l$. Specifically, $a_{lk}(t) = 1$ indicates that user $k$ is associated with BS $l$, and $0$ otherwise. Consequently, the overall action profile for all BSs at time slot $t$ is described by $\boldsymbol{a}(t) = \left[\boldsymbol{a}_{1}(t), \ldots, \boldsymbol{a}_{L}(t)\right]$.

\item 
Reward: The reward function of each BS reflects the performance gain achieved by a chosen action. Specifically, the reward associated with group $o$ of BS $l$ at time slot $t$ is defined as $c_{lo}(t)$, which represents the achievable data rate for group $o$ served by BS $l$ under the action $\boldsymbol{a}_{l}(t)$.

Next, we formally define the total reward across all BSs. Each BS $l$ is limited to its local partial state $\boldsymbol{\xi}_{lt}$ at time slot $t$ and is unaware of the association decisions of neighboring BSs. This limitation can result in a user being scheduled by multiple BSs simultaneously, which is practically undesirable. To address this issue, a penalty term $\rho < 0$ is introduced into the reward function to discourage such conflicting associations. Thus, the total reward for all BSs is defined as follows

\begin{equation}\label{rewardfunction}
\begin{split}
&r\left( \boldsymbol\xi _{t}, \boldsymbol{a}\left(t\right)    \right) \\ &=
\sum_{l=1 }^{L} r_{l}\left( \boldsymbol\xi _{lt}, \boldsymbol{a}_{l}\left(t\right)    \right)  \\
&=\sum_{l=1}^{L}{\sum_{o\in{\mathcal O}}}\left[\left(1-\mathbbm{1}_{\left\{\sum_{\zeta \ne l,\zeta\in{\mathcal L}}a_{\zeta k}\left(t\right)=0\right\}}\right)\rho
+c_{l{o}}\left(t\right)\right],
\end{split}
\end{equation}
where $r_{l}\left( \boldsymbol\xi _{lt}, \boldsymbol{a}_{l}\left(t\right)    \right)={\sum_{o\in{\mathcal O}}}\left[\left(1-\mathbbm{1}_{\left\{\sum_{\zeta \ne l,\zeta\in{\mathcal L}}a_{\zeta k}\left(t\right)=0\right\}}\right)\rho
+c_{l{o}}\left(t\right)\right]$ denotes the reward of BS $l$ at time slot $t$ and $\mathbbm{1}_{\left\{\sum_{\zeta \ne l,\zeta\in{\mathcal L}}a_{\zeta k}\left(t\right)=0\right\}}$ is an indicator function that denotes whether user $k$ is associated with other BSs. Specifically, the indicator function $\mathbbm{1}_{\left\{\sum_{\zeta \ne l,\zeta\in{\mathcal L}}a_{\zeta k}\left(t\right)=0\right\}}$ evaluates to $0$ if user $k$ is associated with other BSs, and $1$ otherwise.  It can be observed from~(\ref{rewardfunction}) that whenever a user is simultaneously scheduled by multiple BSs, the penalty term $\rho$ is incurred in the reward.

\item 
Individual $Q$ value function: For each BS $l$, the individual $Q$ value function $Q_{\boldsymbol{\vartheta}_{lt}}(\boldsymbol{\xi}_{lt}, \boldsymbol{a}_{l}(t))$ quantifies the expected  reward given the local partial state $\boldsymbol{\xi}_{lt}$ of BS $l$ and the executed action $\boldsymbol{a}_{l}(t)$ at time slot $t$. Specifically, the proposed scheme adopts a deep Q-network (DQN), wherein a deep neural network (DNN) parameterized by $\boldsymbol{\vartheta}_{lt}$ is utilized to approximate the individual $Q$ value function at time slot $t$. 
Due to the inherent partial observability, each BS is restricted to acquiring state information solely from users within its local coverage area. Consequently, the BSs must exchange their individual $Q$  values to cooperatively approximate the global $Q$ value function, the mechanism of which is elaborated in the following bullet.

\item 
Global $Q$ value function: The global $Q$ value function, denoted by $Q_{tot}(\boldsymbol{\xi}_{t}, \boldsymbol{a}(t))$, represents the expected total reward across all BSs. Under the VD-RL framework, the global $Q$ value is  decomposed into the sum of the individual $Q$ values from all BSs, which is  formulated as \cite{VDN}
\begin{equation}\label{TD error}
Q_{tot} \left( \boldsymbol\xi _{t}, \boldsymbol{a}\left(t\right) \right)= \sum_{l=1 }^{L} Q_{\boldsymbol{\vartheta}_{lt}} \left( \boldsymbol\xi _{lt}, \boldsymbol{a}_{l}\left(t\right) \right).
\end{equation}
Each BS $l$ aims to cooperatively maximize the overall expected  reward. Upon convergence of the training process, each BS leverages the learned $Q$ function to determine the  BS-user associations and hybrid beamforming matrices, thereby maximizing the aggregate sum rate across all multicast groups. 

\end{itemize}

It should be noted that the additive decomposition
in (\ref{TD error}) is a value-function approximation adopted
by the proposed VD-RL algorithm, rather than an
exact factorization of the true global action-value
function. Although the reward function in (\ref{rewardfunction})
introduces coupling among BSs, the individual
Q-value functions are cooperatively trained
using the global reward and a common
temporal-difference error. Therefore, the
additive decomposition does not require
independent reward functions for individual BSs.

 \subsubsection{VD-RL
Solution}

First, each agent acquires its local observation, consisting of the partial state $\boldsymbol{\xi}_{lt}$ and the executed action $\boldsymbol{a}_l(t)$ at time slot $t$. The agents then exchange these local observations to compute their individual rewards $r_{l}(\boldsymbol{\xi}_{lt}, \boldsymbol{a}_{l}(t))$ and the overall reward $r(\boldsymbol{\xi}_{t}, \boldsymbol{a}(t)) = \sum_{l=1}^{L} r_{l}(\boldsymbol{\xi}_{lt}, \boldsymbol{a}_{l}(t))$. Finally, each agent updates its individual $Q$ value function by minimizing the global $Q$ value loss. Specifically, the loss function associated with the global $Q$ value function $Q_{tot}(\boldsymbol{\xi}_{t}, \boldsymbol{a}(t))$ is defined as
\begin{equation}\label{Qfunction}
\begin{split}
A\left( \boldsymbol{\vartheta}_{1t}, \ldots,  \boldsymbol{\vartheta}_{Lt} \right)  =&\mathbb{E}\left[ Q_{tot} \left( \boldsymbol\xi _{t}, \boldsymbol{a}\left(t\right) \right) - r\left( \boldsymbol\xi _{t}, \boldsymbol{a}\left(t\right)    \right) \right. \\ &    - \max\limits_{\boldsymbol{a}\left(t+1\right)'} Q_{tot} \left( \boldsymbol\xi _{t+1}, {\boldsymbol{a}\left(t+1\right)'} \right)  \left. \right ]^2, 
\end{split}
\end{equation}
where $\max\limits_{\boldsymbol{a}\left(t+1\right)'} Q_{tot} \left( \boldsymbol\xi _{t+1}, {\boldsymbol{a}\left(t+1\right)'} \right)$ denotes the maximum global $Q$ value at the  state $\boldsymbol\xi _{t+1}$.
Since the global $Q$ value is decomposed as the sum of individual $Q$ values, a monotonic improvement in each individual $Q$ value directly translates to an improvement in the global $Q$ value. In other words, an action yielding a high individual $Q$ value for a particular agent is also beneficial to the overall wireless network performance. Consequently, the training objective of each BS reduces to optimizing its own individual $Q$ value function, which in turn drives the maximization of the global $Q$ value. The individual $Q$ value function $Q_{\boldsymbol{\vartheta}_{lt}}(\boldsymbol{\xi}_{lt}, \boldsymbol{a}_{l}(t))$ of each BS is updated via gradient descent as

\begin{equation}\label{revisedQ}
\boldsymbol{\vartheta}_{lt}\leftarrow \boldsymbol{\vartheta}_{lt}-\lambda_{\boldsymbol{\vartheta}_{lt}} \nabla_{\boldsymbol{\vartheta}_{lt}} A\left( \boldsymbol{\vartheta}_{1t}, \ldots,  \boldsymbol{\vartheta}_{Lt} \right),
\end{equation}
where $\lambda_{\boldsymbol{\vartheta}_{lt}}$ is the updating rate and $\nabla_{\boldsymbol{\vartheta}_{lt}} A\left( \boldsymbol{\vartheta}_{1t}, \ldots,  \boldsymbol{\vartheta}_{Lt} \right)$ is the gradient of the global $Q$ value function, which can be mathematically expressed as
\begin{equation}\label{gradient}
\begin{split}
&\nabla_{\boldsymbol{\vartheta}_{lt}} A\left( \boldsymbol{\vartheta}_{1t}, \ldots,  \boldsymbol{\vartheta}_{Lt} \right)\\ &=\nabla_{\boldsymbol{\vartheta}_{lt}} \left[ Q_{tot} \left( \boldsymbol\xi _{t}, \boldsymbol{a}\left(t\right) \right) - r\left( \boldsymbol\xi _{t}, \boldsymbol{a}\left(t\right)    \right)  \right. \\ &  \quad \ - \max\limits_{\boldsymbol{a}\left(t+1\right)'} Q_{tot} \left( \boldsymbol\xi _{t+1}, {\boldsymbol{a}\left(t+1\right)'} \right)  \left. \right]^2\\
&=2\Delta Q_{tot} \nabla_{\boldsymbol{\vartheta}_{lt}} Q_{\boldsymbol{\vartheta}_{lt}} \left( \boldsymbol\xi _{lt}, \boldsymbol{a}_{l}\left(t\right) \right),
\end{split}
\end{equation}
where $\Delta Q_{tot}= Q_{tot} \left( \boldsymbol\xi _{t}, \boldsymbol{a}\left(t\right) \right) - r\left( \boldsymbol\xi _{t}, \boldsymbol{a}\left(t\right)    \right) -\max\limits_{\boldsymbol{a}\left(t+1\right)'} Q_{tot} \left( \boldsymbol\xi _{t+1}, {\boldsymbol{a}\left(t+1\right)'} \right)$.

\subsection{Optimization of hybrid beamforming matrices}

 Upon BS $l$ executing action $\boldsymbol{a}_{l}(t)$, problem~(\ref{PF1}) reduces to
\begin{subequations}   \label{simplified}
	\begin{align}
	 &\mathop {\max}_{\boldsymbol{F}_{l}^\textrm{R}\left(t\right), \boldsymbol{F}^\textrm{B}_{{l,o}}\left(t\right), \boldsymbol{W}_{kl}^{\textrm{R}}\left(t\right), \boldsymbol{W}_{kl}^{\textrm{B}}\left(t\right) }     \mathbb E \left(\sum\limits_{l = 1}^L \sum\limits_{o = 1}^{O}  { c_{lo}\left(t\right) }\right) \tag{\theequation} \\
  	 & \textrm{s.t.}	 
	  \ \ \sum\limits_{o = 1}^{O} \left\| \boldsymbol{F}_{l,o}^\textrm{R}\left(t\right)\boldsymbol{F}_{l,o}^\textrm{B}\left(t\right)\right\|_F^2 \le D,  \forall l,\\
    & \ \ \, \quad {\rm{      }}\left| \boldsymbol{F}_{l}^\textrm{R}\left(i,j\right)\left(t\right) \right| = \left| \boldsymbol{W}_{kl}^{\textrm{R}}\left(i,j\right)\left(t\right) \right| = 1, \forall i,j.
	\end{align}
\end{subequations}

To optimize $\boldsymbol{F}_{l}^\textrm{R}\left(t\right), \boldsymbol{F}^\textrm{B}_{{l,o}}\left(t\right), \boldsymbol{W}_{kl}^{\textrm{R}}\left(t\right)$, and $\boldsymbol{W}_{kl}^{\textrm{B}}\left(t\right)$, we invoke the method developed in our prior work \cite{10196368}.
We first need to simplify the optimization problem in (\ref{simplified}).
$\boldsymbol{F}_{l,o}^\textrm{R}\left(t\right)\boldsymbol{F}^\textrm{B}_{{l,o}}\left(t\right)$, involving baseband digital and analog beamforming, denotes the transmit beamforming matrix for group $o$ at BS $l$. $\boldsymbol{W}_{kl}^{\textrm{R}}\left(t\right) \boldsymbol{W}_{kl}^{\textrm{B}}\left(t\right)$, involving baseband digital and analog combiners, denotes the effective receive beamforming matrix of user $k$ at BS $l$. 
Hence, by treating $\boldsymbol{F}_{l,o}^{\textrm{R}}(t)\boldsymbol{F}_{l,o}^{\textrm{B}}(t)$ and $\boldsymbol{W}_{kl}^{\textrm{R}}(t)\boldsymbol{W}_{kl}^{\textrm{B}}(t)$ as integrated matrices, the hybrid beamforming problem in (\ref{simplified}) can be relaxed into a fully digital beamforming problem. Specifically, we define $\boldsymbol{B}_{l,o}(t)=\boldsymbol{F}_{l,o}^{\textrm{R}}(t)\boldsymbol{F}_{l,o}^{\textrm{B}}(t)\in\mathbb{C}^{N^\textrm{B}\times\zeta}$ and $\boldsymbol{J}_{kl}(t)=\boldsymbol{W}_{kl}^{\textrm{R}}(t)\boldsymbol{W}_{kl}^{\textrm{B}}(t)\in\mathbb{C}^{N^\textrm{U}\times\zeta}$ as the equivalent transmit and receive beamforming matrices, respectively, combining the corresponding analog RF and digital baseband components.  Directly solving the hybrid beamforming problem in (\ref{simplified}) is challenging due to the coupling between the analog RF and digital baseband beamforming matrices, as well as the non-convex constant-modulus constraints imposed on the RF beamformers. By treating the analog and digital components as integrated matrices, these constant-modulus constraints can be temporarily relaxed, thereby avoiding the direct optimization of the coupled beamforming matrices. The resulting fully digital beamforming problem allows the BD method to be employed for constructing the transmit and receive beamformers. Subsequently, the FO algorithm approximates these fully digital matrices using feasible analog RF and digital baseband beamforming matrices to obtain a hybrid beamforming solution satisfying the hardware constraints. By substituting $\boldsymbol{B}_{l,o}(t)$ and $\boldsymbol{J}_{kl}(t)$ into (\ref{simplified}) and relaxing the RF constant-modulus constraints, we obtain the following fully digital beamforming problem

\begin{subequations}\label{Transformation}
\begin{align}
& \mathop {\max }_{\boldsymbol{B}_{l,o}\left(t\right), \boldsymbol{J}_{kl}\left(t\right) }    \mathbb E \left(\sum\limits_{l = 1}^L \sum\limits_{o = 1}^{O} { c_{lo}\left(t\right) }\right) \tag{\theequation} \\
& \textrm{s.t.} \ \
 \sum\limits_{o = 1}^{O} \left\| \boldsymbol{B}_{l,o}\left(t\right)\right\|_F^2 \le D,  \forall l.
\end{align}
\end{subequations}
Constraint (\ref{simplified}b) imposes element-wise constant-modulus restrictions on both the transmit and receive RF beamforming matrices. In the fully digital design stage, \(\boldsymbol{B}_{l,o}(t)\) and \(\boldsymbol{J}_{kl}(t)\) are treated as design variables without requiring a factorization into RF and baseband components satisfying the hybrid hardware constraints. Accordingly, constraint (\ref{simplified}b) is temporarily relaxed, whereas the total BS transmit-power constraint is retained. Thus, problem (\ref{Transformation}) is a fully digital relaxation of problem (\ref{simplified}).
 To optimize $\boldsymbol{B}_{l,o}\left(t\right)$ and $\boldsymbol{J}_{kl}\left(t\right)$ in problem (\ref{Transformation}), 
 $\tilde{\boldsymbol{H}}_{lo}$ is first introduced as
	\begin{equation}\label{PF3}
	\tilde{\boldsymbol{H}}_{lo} \buildrel \Delta \over = \left [ \bar{\boldsymbol{H}}_{11},\ldots,\bar{\boldsymbol{H}}_{l{\left (o-1\right )}},\bar{\boldsymbol{H}}_{l{\left (o+1\right )}},\ldots,\bar{\boldsymbol{H}}_{L{N_L}} \right ]^{\textrm{T}},
	\end{equation}
where $\boldsymbol\phi _{lk}\left(t\right)$ is omitted without loss of generality, $N_L$ represents the index of the maximum group under BS $L$ 
 and  $\bar{\boldsymbol{H}}_{l{\left (o-1\right )}} = \left [ {\boldsymbol{H}}_{l{\left (o-1\right )},1}, {\boldsymbol{H}}_{l{\left (o-1\right )},2},\ldots,{\boldsymbol{H}}_{l{\left (o-1\right )},E_{l{\left (o-1\right )}}} \right ] $ represents the matrix corresponding to the effective channels for all members in group $o$-1 of BS $l$ with $E_{l{\left (o-1\right )}}$ denoting the number of members of group $o$-1.  Denoting the rank of $\tilde{\boldsymbol{H}}_{lo}$ as $\tilde{L}_{k}$, we next detail the application of the BD method to derive the transmit precoders and receive combiners. To eliminate the interference between groups, 
$\tilde{\boldsymbol{V}}_{lo}^{\left(0\right)} \in \mathbb{C}^{N^\textrm{B} \times \left({N^\textrm{B}-\tilde{L}_{k}}\right)}$ is defined as
 
\begin{equation}\label{interfer}
	\tilde{\boldsymbol{V}}_{lo}^{\left(0\right)}=\text{null} \left(\tilde{\boldsymbol{H}}_{lo}\right),
	\end{equation}
where $\text{null} \left(\tilde{\boldsymbol{H}}_{lo}\right)$ indicates that $\tilde{\boldsymbol{V}}_{lo}^{\left(0\right)}$ is in the $\text{null}$ space of $\tilde{\boldsymbol{H}}_{lo}$. Thus, it follows that $\tilde{\boldsymbol{H}}_{lo} \tilde{\boldsymbol{V}}_{lo}^{\left(0\right)} = \boldsymbol 0$.  Since $\tilde{\boldsymbol{H}}_{lo}$ has $N^\textrm{B}$ columns and rank $\tilde{L}_{k}$, the dimension of its null space is $N^\textrm{B}-\tilde{L}_{k}$. Therefore, to support \(\zeta\) spatial data streams while satisfying the inter-group interference-nulling constraints, the BD design requires
$N^\textrm{B}-\tilde{L}_{k}\geq\zeta$.
This condition ensures that at least \(\zeta\) linearly independent transmit directions are available in the null space of $\tilde{\boldsymbol{H}}_{lo}$.
 To perfectly diagonalize and eliminate the inter-stream interference for each user, we perform the SVD on the effective channel ${\boldsymbol{H}}_{lo,k} \tilde{\boldsymbol{V}}_{lo}^{\left(0\right)}$, which is expressed as
	\begin{equation}\label{muitiplrstreams}
	{\boldsymbol{H}}_{lo,k} \tilde{\boldsymbol{V}}_{lo}^{\left(0\right)}= \left [ {\boldsymbol{U}}_{k}^{\left(1\right)} , {\boldsymbol{U}}_{k}^{\left(2\right)} \right ] \left [ \begin{matrix}  {\boldsymbol{\Sigma}}_{k}^{\left(1\right)}&\boldsymbol{0}&\\\boldsymbol{0}&{\boldsymbol{\Sigma}}_{k}^{\left(2\right)}& \end{matrix} \!\!\!\!\! \right ] \left [ {\boldsymbol{V}}_{k}^{\left(1\right)} , {\boldsymbol{V}}_{k}^{\left(2\right)} \right ]^{\textrm{H}},
	\end{equation}
where ${\boldsymbol{H}}_{lo,k}$ represents the effective channels for user $k$ in group $o$ of BS $l$. The rank of the matrix ${\boldsymbol{H}}_{lo,k} \tilde{\boldsymbol{V}}_{lo}^{\left(0\right)}$ is denoted by ${L}_{k}$. The column vectors of $\boldsymbol{U}_{k}^{\left(1\right)} \in \mathbb{C}^{N^\textrm{U} \times \zeta}$, $\boldsymbol{U}_{k}^{\left(2\right)} \in \mathbb{C}^{N^\textrm{U} \times \left({{L}_{k}-\zeta}\right)}$, $\boldsymbol{V}_{k}^{\left(1\right)} \in \mathbb{C}^{\left({N^\textrm{B}-\tilde{L}_{k}}\right) \times \zeta}$, and $\boldsymbol{V}_{k}^{\left(2\right)} \in \mathbb{C}^{\left({N^\textrm{B}-\tilde{L}_{k}}\right) \times \left({{L}_{k}-\zeta}\right)}$ can form orthonormal sets.  ${\boldsymbol{\Sigma}}_{k}^{\left(1\right)}\in  \mathbb{C}^{\zeta \times \zeta}$ and ${\boldsymbol{\Sigma}}_{k}^{\left(2\right)}\in  \mathbb{C}^{ \left({{L}_{k}-\zeta}\right) \times \left({{L}_{k}-\zeta}\right) }$ are defined as diagonal matrices composed of singular values.   In addition to the null-space dimension condition, the projected effective channel of each served user should satisfy
$
L_k=\text{rank}\!\left(
{\boldsymbol{H}}_{lo,k} \tilde{\boldsymbol{V}}_{lo}^{\left(0\right)}
\right)\geq\zeta$,
such that at least \(\zeta\) nonzero singular values are available for supporting the \(\zeta\) spatial data streams. This condition also requires \(N^{\mathrm U}\geq\zeta\).    For a fixed BS--group pair $\left(l,o\right)$ and time slot $t$,
index the $E_{lo}$ users in the multicast group by
$k=1,\ldots,E_{lo}$. We construct the aggregated beamforming
matrix as
\begin{equation} \label{abm1}
    \begin{split}
        \boldsymbol{C}_{lo}
        =\widetilde{\boldsymbol{V}}_{lo}^{\left(0\right)}
          \sum_{k=1}^{E_{lo}}\boldsymbol{V}_{k}^{\left(1\right)} 
        =\left[
            \boldsymbol{c}_{lo,1},\ldots,\boldsymbol{c}_{lo,\zeta}
          \right].
    \end{split}
\end{equation}
The singular vectors corresponding to different users need not be
mutually orthogonal. 
Each nonzero column is normalized as
\begin{equation}\label{abm2}
    \begin{split}
        \widehat{\boldsymbol{c}}_{lo,i}
        =\frac{\boldsymbol{c}_{lo,i}}
          {\left\|\boldsymbol{c}_{lo,i}\right\|_{2}},
          \qquad i=1,\ldots,\zeta, 
    \end{split}
\end{equation}
\begin{equation}\label{abm66}
    \begin{split}
 \widehat{\boldsymbol{C}}_{lo}
        =\left[
            \widehat{\boldsymbol{c}}_{lo,1},\ldots,
            \widehat{\boldsymbol{c}}_{lo,\zeta}
          \right].
    \end{split}
\end{equation}
This construction requires
$\left\|\boldsymbol{c}_{lo,i}\right\|_{2}>0$.
The fully digital transmit beamforming matrix is then given by
\begin{equation}\label{eq:normalized_multicast_precoder}
    \begin{split}
        \boldsymbol{B}_{l,o}\left(t\right)
        &=\widehat{\boldsymbol{C}}_{lo}
          \boldsymbol{G}_{lo}^{1/2},
    \end{split}
\end{equation}
where
$\boldsymbol{G}_{lo}=\operatorname{diag}\left(d_1,\ldots,d_{\zeta}\right)$
is the power allocation matrix, with $d_i\geq 0$.
To  nullify the inter-stream interference within group $o$, the fully digital receive combiner $\boldsymbol{J}_{kl}\left(t\right)$ for the $k$-th user at BS $l$ is given by
	\begin{equation}\label{PF7+}
	\begin{split}
	\boldsymbol{J}_{kl}\left(t\right)&={\boldsymbol{U}}_k^{\left(1\right)}.
	\end{split}
	\end{equation}

Substituting the equal power allocation
$d_i=D/\left(N_l\zeta\right)$ into
(\ref{eq:normalized_multicast_precoder}) yields
\begin{equation}\label{eq:normalized_multicast_equal_power}
    \begin{split}
        \boldsymbol{B}_{l,o}\left(t\right)
        &=\sqrt{\frac{D}{N_l\zeta}}
          \widehat{\boldsymbol{C}}_{lo}.
    \end{split}
\end{equation}
Equivalently, the fully digital transmit precoder can be written as
\begin{equation}\label{eq:normalized_multicast_expanded}
    \begin{split}
        \boldsymbol{B}_{l,o}\left(t\right)
        &=\sqrt{\frac{D}{N_l\zeta}}
          \left[
            \frac{\boldsymbol{c}_{lo,1}}
                 {\left\|\boldsymbol{c}_{lo,1}\right\|_{2}},
            \ldots,
            \frac{\boldsymbol{c}_{lo,\zeta}}
                 {\left\|\boldsymbol{c}_{lo,\zeta}\right\|_{2}}
          \right].
    \end{split}
\end{equation}
  
 The fully digital receive combiner $\boldsymbol{J}_{kl}\left(t\right)$ for user $k$ at BS $l$ is given by
	\begin{equation}\label{PF7+}
	\begin{split}
	\boldsymbol{J}_{kl}\left(t\right)&={\boldsymbol{U}}_k^{\left(1\right)}.
	\end{split}
	\end{equation}

Then, given $\boldsymbol{B}_{l,o}\left(t\right)$ and $\boldsymbol{J}_{kl}\left(t\right)$,  we  can use the algorithm in \cite{8646553} to optimize $\boldsymbol{F}_{l}^\textrm{R}\left(t\right), \boldsymbol{F}^\textrm{B}_{{l,o}}\left(t\right), \boldsymbol{W}_{kl}^{\textrm{R}}\left(t\right)$ and $\boldsymbol{W}_{kl}^{\textrm{B}}\left(t\right)$.
As a result, the proposed algorithm for solving problem (\ref{PF1}) is summarized in
Algorithm 1.

\begin{algorithm}[t]
\caption{Proposed  Algorithm for Solving  Problem (\ref{PF1})}
\begin{algorithmic}[1]
\STATE Initialize the DNN parameters $\boldsymbol{\vartheta}_{lt}$ of each BS $l$ at time slot $t$.  
\FOR {each iteration}
\FOR {each BS $l$}
\FOR {each time slot $t$}
\STATE  Acquire the local observation $ \boldsymbol\xi _{lt}$ of the environment.
\STATE  Determine the action according to the $\epsilon$-greedy policy.
\STATE Obtain $\boldsymbol{B}_{l,o}\left(t\right)$ and $\boldsymbol{J}_{kl}\left(t\right)$ by BD method.
\STATE Optimize $\boldsymbol{F}_{l}^\textrm{R}\left(t\right), \boldsymbol{F}^\textrm{B}_{{l,o}}\left(t\right), \boldsymbol{W}_{kl}^{\textrm{R}}\left(t\right)$ and $\boldsymbol{W}_{kl}^{\textrm{B}}\left(t\right)$ by a fast optimization algorithm.
\STATE Compute the reward $r_{l}\left( \boldsymbol\xi _{lt}, \boldsymbol{a}_{l}\left(t\right)    \right)$, the $Q$ function values $Q_{\boldsymbol{\vartheta}_{lt}} \left( \boldsymbol\xi _{lt}, \boldsymbol{a}_{l}\left(t\right) \right)$, and $\max\limits_{\boldsymbol{a}_{l}\left(t+1\right)'} Q_{\boldsymbol{\vartheta}_{l\left(t+1\right)}} \left( \boldsymbol\xi _{l\left(t+1\right)}, {\boldsymbol{a}_{l}\left(t+1\right)'} \right)$.
\ENDFOR
\STATE Share the rewards and $Q$ function values with  other BSs.
\ENDFOR
\FOR {each BS $l$}
\STATE  Compute $\Delta Q_{tot}$ in (\ref{gradient}) using the shared rewards and $Q$-values.
\STATE Use $\Delta Q_{tot}$ to update $\boldsymbol{\vartheta}_{lt}$ based on (\ref{revisedQ}).
\ENDFOR
\ENDFOR
\end{algorithmic}
\label{algorithm_1}
\end{algorithm}

\subsection{Complexity and Action-Selection Property of the Proposed Algorithm}\label{Complexity and Convergence of the Proposed Algorithm}

\subsubsection{Complexity of the Proposed Algorithm}

Next, we explain
the complexity of training the VD-RL algorithm. The
complexity of training the VD-RL scheme primarily lies in
training the distributed DQNs for each BS, which involves
determining BS-user associations, fully digital beamforming
at both BSs and users, as well as hybrid beamforming at both
BSs and users.

Regarding the complexity of training distributed DQNs, each BS is required to independently update its RL parameters $\boldsymbol{\vartheta}_{lt}$. 
  Therefore, the time complexity of training a DQN depends on the input dimension of $\boldsymbol\xi _{lt}$, the output dimension of $\boldsymbol{a}_{l}\left(t\right)$, and the number of neurons within each hidden layer \cite{9691928}. Let ${m}_{y}$ denote the number of neurons in the $y$-th hidden layer, and ${X}_l$ represent the total number of hidden layers of the DQN at BS $l$. Accordingly, the time complexity of training a DQN is  $\mathcal O\left( d\left( \sum\limits_{y = 1}^{{X}_l} {m}_{y} {m}_{y+1} +  \left | \boldsymbol\xi _{l}\right | {m}_{1} + \left | \boldsymbol{a}_{l}\right | {X}_l   \right) \right)$, where $d$ is the number of iterations at which the  VD-RL algorithm converges, $\left | \boldsymbol\xi _{l} \right |$ and $\left | \boldsymbol{a}_{l} \right |$ denote the dimensions of $\boldsymbol\xi _{lt}$ and $\boldsymbol{a}_{l}\left(t\right)$, respectively.

\subsubsection{Action-Selection Property of the Proposed Algorithm}

In this section, we characterize the action-selection property of the
proposed VD-RL scheme. Specifically, under the additive value decomposition
in (\ref{TD error}) and independent candidate action selection across BSs, we show
that the joint greedy action can be obtained from the individual greedy
actions.

To allow the BSs to update their value functions independently, it is necessary to identify an efficient and adaptable way to decompose the global $Q$  value function $Q_{tot} \left( \boldsymbol\xi _{t}, \boldsymbol{a}\left(t\right) \right)$ among all the BSs. A widely recognized requirement is the necessity of meeting the individual-global-max (IGM) condition.

\begin{myDef} \label{defIGM}
For a multiagent system with $L$ agents, the 
IGM condition means that a joint action maximizing the represented
global $Q$ value function can be obtained by combining the actions
maximizing the individual $Q$ value functions. The statement is presented as follows

\begin{equation} \label{IMGSTA}
\begin{split}
\text{arg}  \max\limits_{\boldsymbol{a}\left(t\right)}  Q_{tot} \left( \boldsymbol\xi _{t}, \boldsymbol{a}\left(t\right) \right) =
   \quad \left( \begin{matrix} \text{arg} \max\limits_{\boldsymbol{a}_{1}\left(t\right)}  Q_{\boldsymbol{\vartheta}_{1t}} \left( \boldsymbol\xi _{1t}, \boldsymbol{a}_{1}\left(t\right) \right)&\\ \text{arg} \max\limits_{\boldsymbol{a}_{2}\left(t\right)}  Q_{\boldsymbol{\vartheta}_{2t}} \left( \boldsymbol\xi _{2t}, \boldsymbol{a}_{2}\left(t\right) \right)&\\\vdots&\\\text{arg} \max\limits_{\boldsymbol{a}_{L}\left(t\right)}  Q_{\boldsymbol{\vartheta}_{Lt}} \left( \boldsymbol\xi _{Lt}, \boldsymbol{a}_{L}\left(t\right) \right)\end{matrix} \!\!\!\!\! \right).
\end{split}
\end{equation}
\end{myDef}

When the IGM condition is satisfied, optimizing the global $Q$ value is equivalent to optimizing the individual $Q$ value for each agent.  We can also say that $Q_{tot} \left( \boldsymbol\xi _{t}, \boldsymbol{a}\left(t\right) \right)$ is factorized by $Q_{\boldsymbol{\vartheta}_{lt}} \left( \boldsymbol\xi _{lt}, \boldsymbol{a}_{l}\left(t\right) \right)$, or that $Q_{\boldsymbol{\vartheta}_{lt}} \left( \boldsymbol\xi _{lt}, \boldsymbol{a}_{l}\left(t\right) \right)$ is the factor of $Q_{tot} \left( \boldsymbol\xi _{t}, \boldsymbol{a}\left(t\right) \right)$.  
By Definition \ref{defIGM} of IGM, we can find that  the global $Q$ value function $Q_{tot} \left( \boldsymbol\xi _{t}, \boldsymbol{a}\left(t\right) \right)$ is factorized by individual $Q$ value function $Q_{\boldsymbol{\vartheta}_{lt}} \left( \boldsymbol\xi _{lt}, \boldsymbol{a}_{l}\left(t\right) \right)$.    Then, Lemma \ref{lemmafac} provides the sufficient condition for $Q_{\boldsymbol{\vartheta}_{lt}} \left( \boldsymbol\xi _{lt}, \boldsymbol{a}_{l}\left(t\right) \right)$ to satisfy the IGM condition.

Under the additive decomposition in (\ref{TD error}), the IGM condition in (\ref{IMGSTA})
holds for a candidate joint-action space that is the Cartesian product
of the individual BS action spaces. The following lemma formalizes
this action-selection property.

\begin{lemma} \label{lemmafac}
Suppose that the represented global $Q$ value function satisfies the
additive decomposition in (\ref{TD error}) for any candidate joint action
$\boldsymbol{a}\left(t\right)$, and that the candidate actions of the BSs
can be selected independently. For each BS $l$, let
$\bar{\boldsymbol{a}}_{l}\left(t\right)$ be a greedy action satisfying
$\bar{\boldsymbol{a}}_{l}\left(t\right)\in
\text{arg} \max\limits_{\boldsymbol{a}_{l}\left(t\right)}  Q_{\boldsymbol{\vartheta}_{lt}} \left( \boldsymbol\xi _{lt}, \boldsymbol{a}_{l}\left(t\right) \right)$,
and define
$\bar{\boldsymbol{a}}\left(t\right)= \left[ \bar{\boldsymbol{a}}_{l}\left(t\right) \right]_{l=1 }^{L} $.
Then, the joint action formed by these individual greedy actions satisfies
\begin{equation}\label{eq:r1c2_action_selection}
    \begin{split}
        \bar{\boldsymbol{a}}\left(t\right)
        &\in \text{arg}\max_{\boldsymbol{a}\left(t\right)}
        Q_{tot}\left(\boldsymbol{\xi}_{t},
        \boldsymbol{a}\left(t\right)\right).
    \end{split}
\end{equation}
\end{lemma}

\begin{IEEEproof}
For any candidate joint action $\boldsymbol{a}\left(t\right)$,
we have
\begin{equation}\label{eq:r1c2_action_selection_proof}
    \begin{split}
        &Q_{tot}\left(\boldsymbol{\xi}_{t},
        \bar{\boldsymbol{a}}\left(t\right)\right) \\
        &=\sum_{l=1}^{L}Q_{\boldsymbol{\vartheta}_{lt}}\left(
        \boldsymbol{\xi}_{lt},\bar{\boldsymbol{a}}_{l}\left(t\right)\right)\\
        &\geq \sum_{l=1}^{L}Q_{\boldsymbol{\vartheta}_{lt}}\left(
        \boldsymbol{\xi}_{lt},\boldsymbol{a}_{l}\left(t\right)\right)\\
        &=Q_{tot}\left(\boldsymbol{\xi}_{t},
        \boldsymbol{a}\left(t\right)\right).
    \end{split}
\end{equation}
The equality follows from (\ref{TD error}), and the inequality follows from the
definition of each individual greedy action. Hence,
$\bar{\boldsymbol{a}}\left(t\right)$ maximizes the represented global
$Q$ value function, which establishes the stated action-selection
property. This completes the proof.
\end{IEEEproof}

Lemma \ref{lemmafac} shows that, under the stated conditions, the joint greedy action
with respect to the represented global $Q$ value function can be obtained
by concatenating the individual greedy actions of the BSs. Consequently,
the learned individual $Q$ value functions can be used for decentralized
greedy action selection.

\section{Simulation Results And Analysis}\label{Simulation Results And Analysis}
The simulation setup considers a scenario where five BSs cooperatively serve 50 mobile users. The BSs are positioned at coordinates (50m, 50m, 10m), (200m, 800m, 10m), (400m, 500m, 10m), (750m, 50m, 10m), and (1000m, 600m, 10m). 
The considered simulation area is a rectangle
of $1100$ m $\times$ $900$ m, with horizontal
coordinates in $[0,1100]\times[0,900]$.
Each BS has a horizontal coverage radius of $300$ m.
At the beginning of each episode, the horizontal
user locations are independently and uniformly
sampled from the portion of the simulation area
covered by at least one BS. 
The users follow the discrete-time 2D random
walk model with
$\boldsymbol{p}_{kt}=[0.2,0.2,0.2,0.2,0.2]$,
$v_k=2$ m/s, and $\Delta t=1$ s.
The numbers of antennas
at each BS and each user are
$N^\textrm{B}=128$ and $N^\textrm{U}=64$, respectively.
The total number of propagation paths for each
BS--user link is set to $N_\textrm{p}=7$, including
both LOS and NLOS paths. 
The bandwidth is $W=251.1886$ MHz.
The carrier frequency is $28$ GHz, and the
antenna spacing is $\lambda/2$. For links containing
a LOS component and NLOS scattering paths, the
large-scale path loss follows
$\textrm{PL}(d)=61.4+20\log_{10}(d)+X_\sigma$ dB,
where $d$ is the BS--user distance in meters and
$X_\sigma\sim\mathcal{N}(0,5.8^2)$ represents
shadowing in dB.
Each BS uses an independent feedforward Q-network
with two hidden layers of 64 ReLU units each.
Training uses stochastic gradient descent (SGD) with a learning rate of $10^{-4}$,
a squared one-step temporal difference (TD) loss, and a discount factor
of $\gamma=1$. The replay buffer holds 5000 episodes,
and the minibatch size is 32 episodes.
The exploration probability  decreases
linearly from $1$ to $0.05$ over 30000 training
environment steps and remains constant thereafter.
TD targets are computed using the online network with detached gradients, without employing a separate target network.
The remaining system parameters are summarized in Table~II.   To benchmark the performance of the proposed algorithm, three baseline schemes are considered, which are described as follows:

\begin{table}[t]
	\renewcommand{\arraystretch}{1.4}
	\centering
	\caption{Simulation Parameters}
	\label{table2}
	\begin{tabular}{|c|c|c|c|}
		\hline
		\textbf{Parameters}                    & \textbf{Values} &\textbf{Parameters}                    & \textbf{Values}     \\ \hline
		$L$  & $5$ &  $K$          & 50  \\ \hline
		$N^\textrm{U}$           & 64          &$S^\textrm{U}$            & 4             \\ \hline
		$O$                           &5      &${\zeta}$  & 4            \\ \hline
		$\rho$ & -2           &$D$ & 50 dBm             \\ \hline
		$v_k$                    & 2 m/s   &$\sigma^2$                  & -90 dBm   \\ \hline
                $N^\textrm{B}$
        & 128
        & $N_\textrm{p}$
        & 7 \\ \hline

        $W$
        & 251.1886 MHz
        & $S^\textrm{U}$
        & 20 \\ \hline
	\end{tabular}
\end{table}

\begin{itemize}
\item
a) VDRL-ZF: To address the formulated joint optimization problem, a VD-RL scheme is proposed. The proposed VD-RL algorithm adopts a two-stage optimization strategy, which jointly considers BS-user connections and hybrid beamforming. Specifically, in the first stage, the BS-user connections are determined by VD-RL algorithm. In the second stage,  the hybrid precoders and combiners are designed based on the zero forcing (ZF) scheme with fixed BS-user connections.

\item
b) VDRL-Heuristic: To address the formulated joint optimization problem, a VD-RL scheme is proposed. The proposed VD-RL algorithm adopts a two-stage optimization strategy, which jointly considers BS-user connections and hybrid beamforming. Specifically, in the first stage, the BS-user connections are determined by VD-RL algorithm. In the second stage,  the hybrid precoders and combiners are designed based on a heuristic algorithm with fixed BS-user connections.

\item
c) IDRL-BDFO: To address the formulated joint optimization problem, a independent distributed reinforcement learning (IDRL) algorithm is proposed. The proposed IDRL algorithm adopts a two-stage optimization strategy, which jointly considers BS-user connections and hybrid beamforming. Specifically, in the first stage, the BS-user connections are determined by IDRL algorithm. In the second stage,  the hybrid precoders and combiners are designed based on a BD method and a FO algorithm with fixed BS-user connections.

\end{itemize}

\begin{figure}[!t]
	\centering
	\includegraphics[width=9cm]{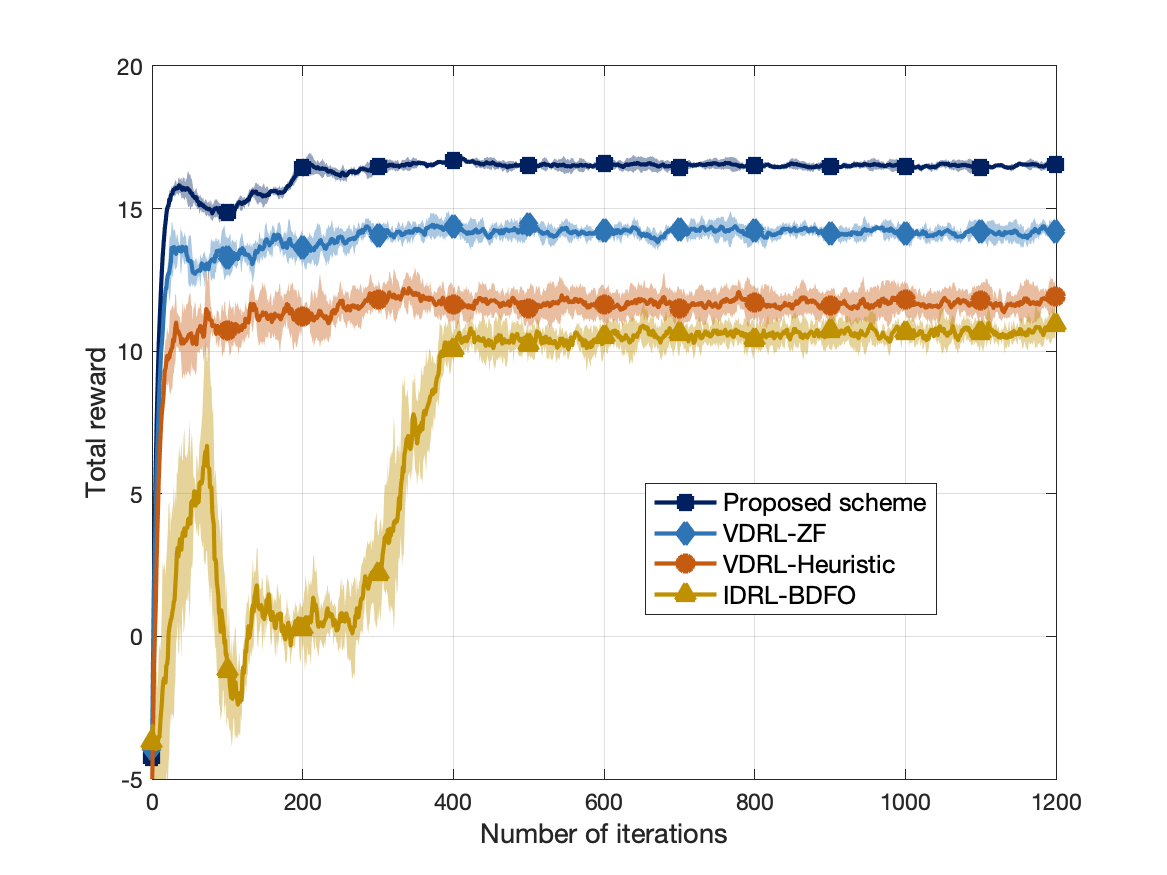}\\
	\caption{Value of the total rewards as the number of iterations.} \label{reward}
\end{figure}

Fig.~\ref{reward} depicts the empirical learning behavior of all
considered schemes. The solid curves represent the mean total rewards
across independent training runs, while the shaded regions indicate
the corresponding 95\% confidence intervals. It can be observed that
the proposed algorithm achieves a rapid increase in the mean total
reward during the initial training stage and gradually stabilizes
after approximately 200 iterations. In contrast, VDRL-ZF and
VDRL-Heuristic exhibit relatively slower learning behavior, while
IDRL-BDFO experiences significant reward fluctuations during the
initial training stage and reaches a relatively stable reward level
after approximately 400 iterations. The relatively rapid empirical
stabilization of the proposed algorithm can be attributed to the
cooperative decision-making mechanism, which enables all BSs to
jointly optimize the sum rate rather than acting independently.
Furthermore, Fig.~\ref{reward} illustrates that the proposed algorithm
outperforms the VDRL-ZF, VDRL-Heuristic, and IDRL-BDFO schemes in terms
of the mean total reward. The performance advantage over the former
two schemes can be attributed to the BD and FO methods, which
effectively mitigate both inter-group and inter-stream interference
without severely sacrificing the desired signal power. In contrast,
VDRL-ZF suffers from significant signal attenuation due to strict
zero-forcing constraints in multicast scenarios, while VDRL-Heuristic
leads to severe interference leakage. The performance improvement
over IDRL-BDFO can be attributed to the cooperative learning mechanism,
which enables all BSs to jointly optimize the global reward rather
than independently maximizing their individual rewards. Moreover,
the relatively narrow confidence bands of the proposed algorithm
indicate lower uncertainty in the estimated mean total reward under
the considered simulation settings.

	

\begin{figure}[!t]
	\centering
	\includegraphics[width=9cm]{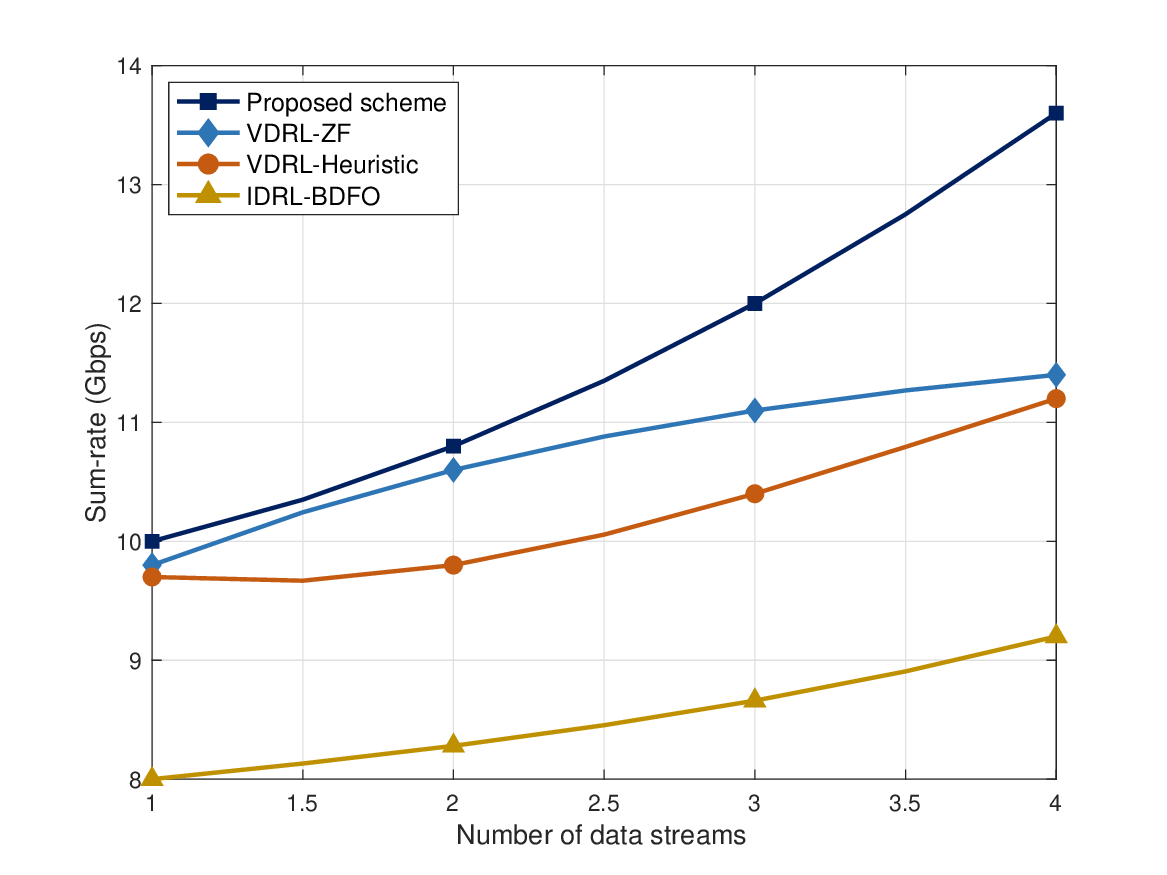}\\
	\caption{Sum-rate versus number of data streams.}\label{RateDS}
\end{figure}

Fig.~\ref{RateDS} illustrates the sum rate of  multicast groups versus the number of data streams. As shown, the performance advantage of the proposed scheme over VDRL-ZF and VDRL-Heuristic becomes increasingly significant with a growing number of data streams. The underlying reason is that when a single data stream is transmitted per user, only inter-group interference exists among users, while intra-user multi-stream interference is nonexistent. As the number of data streams escalates, however, the intra-user multi-stream interference grows proportionally, further degrading the performance of VDRL-ZF and VDRL-Heuristic. The proposed algorithm effectively mitigates both inter-group and inter-stream interference for each user without severely sacrificing the desired signal power. In contrast, VDRL-ZF severely depletes the limited spatial degrees of freedom to forcefully eliminate the escalating interference, resulting in significant signal attenuation and a flattened sum-rate growth. Meanwhile, VDRL-Heuristic suffers from severe interference leakage due to the lack of rigorous interference management.

\begin{figure}

\centering
\subfigure[]{\label{fig:fft:a}

\begin{minipage}[c]{0.5\textwidth}

\centering

\includegraphics[height=5.5cm,width=7.5cm]{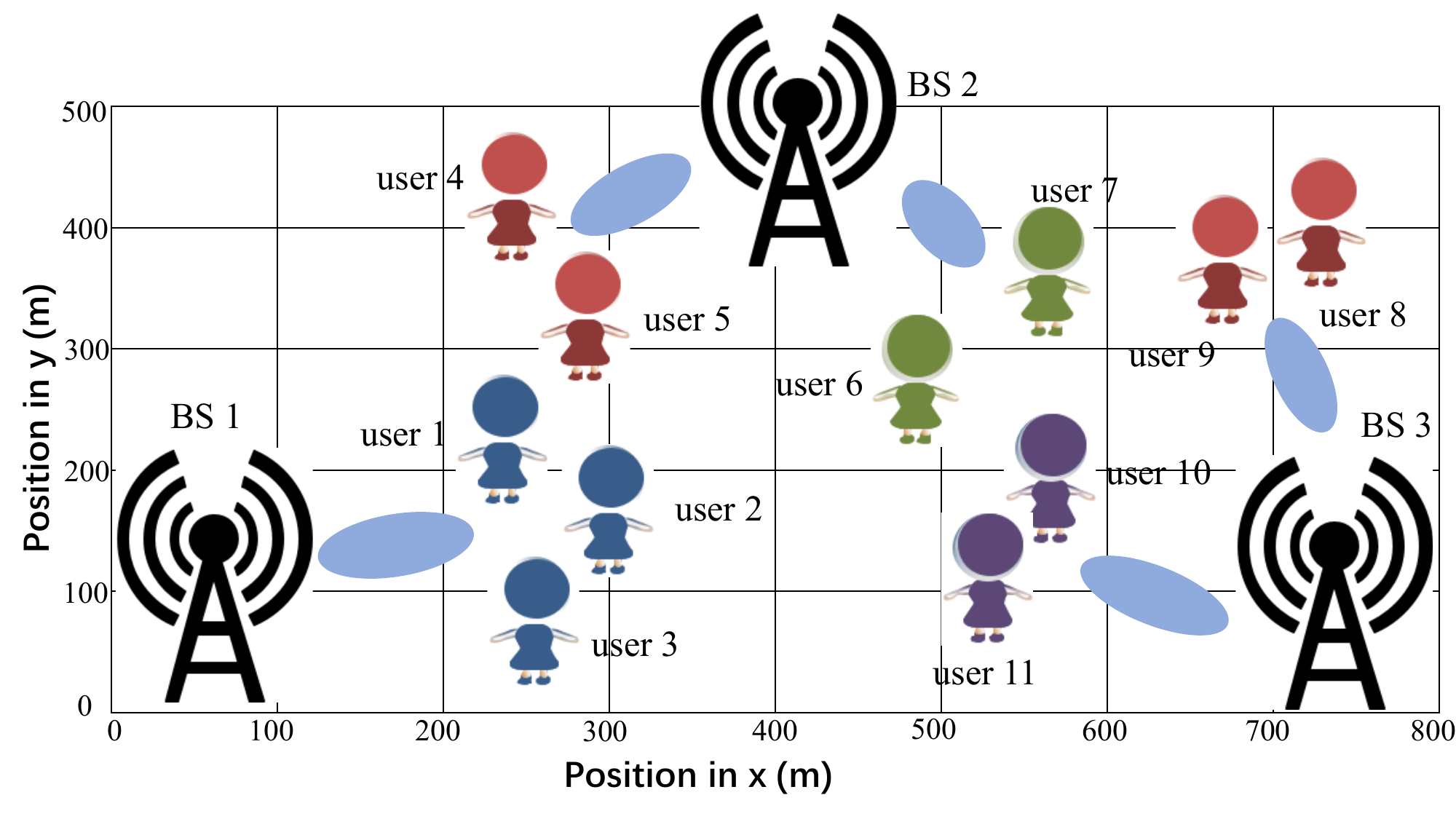}

\end{minipage}%
}

\subfigure[]{\label{fig:fft:a}
\begin{minipage}[c]{0.5\textwidth}

\centering

\includegraphics[height=5.5cm,width=7.5cm]{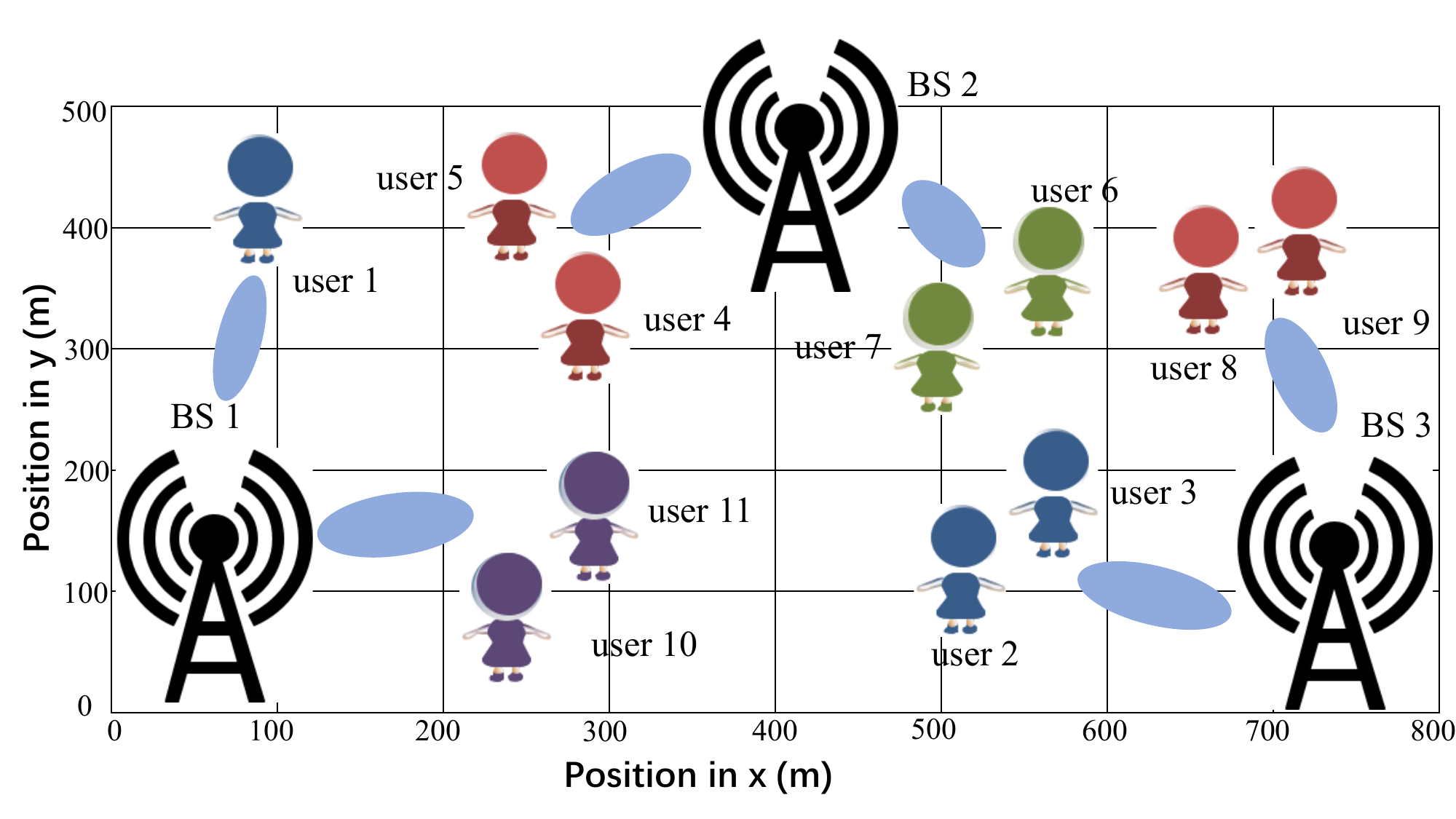}

\end{minipage}
}

\subfigure[]{\label{fig:fft:a}
\begin{minipage}[c]{0.5\textwidth}

\centering

\includegraphics[height=5.5cm,width=7.5cm]{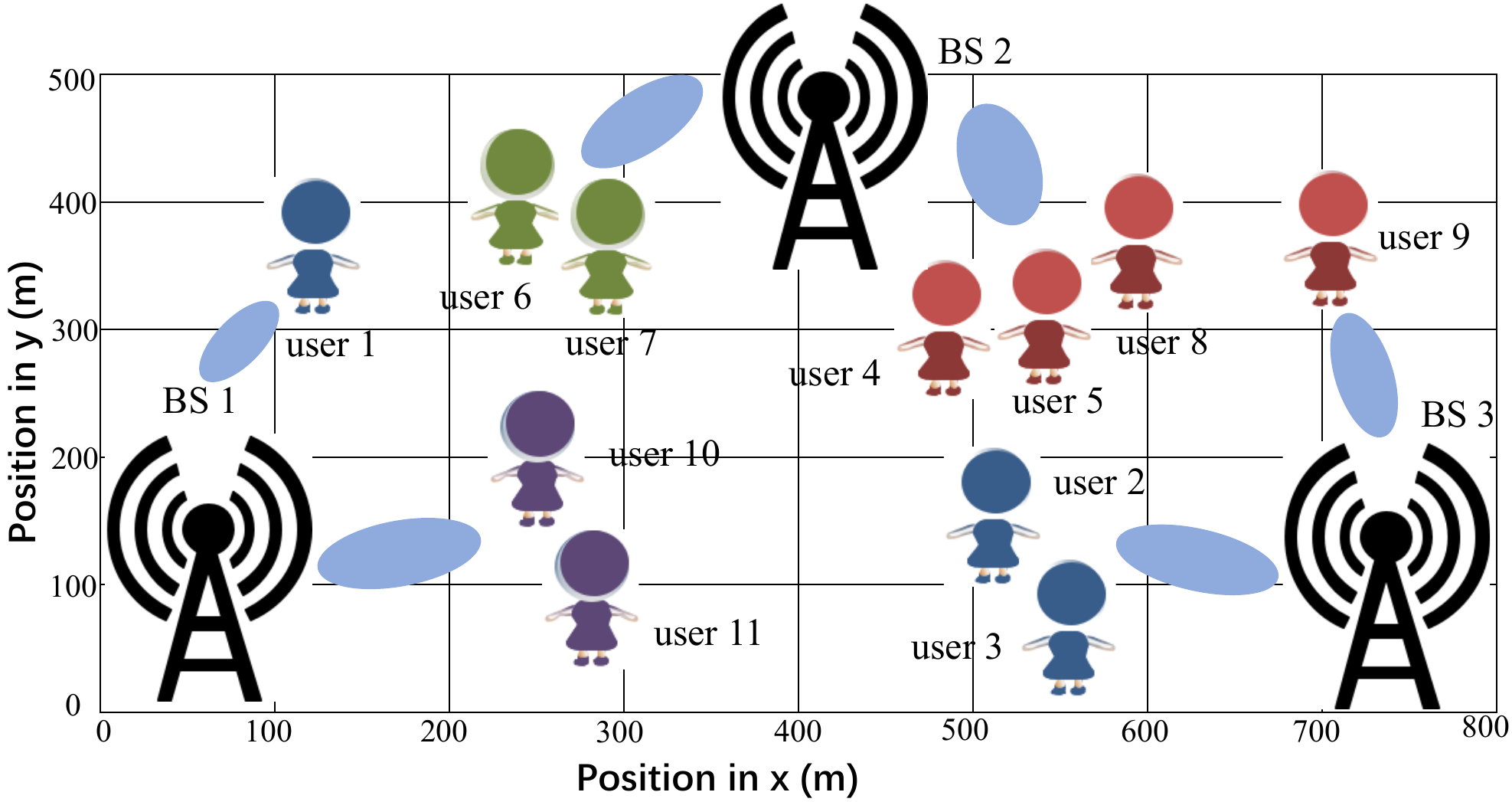}

\end{minipage}
}
\caption{Visualization of BS-user connections and user grouping obtained
by the proposed method.}
\label{Visualization} 
\end{figure}

Fig. \ref{Visualization} presents the user associations under the proposed scheme, where users requesting the same content are denoted by identical colors. As shown in Fig. \ref{Visualization}(a), geographically proximate users with identical content requests are highly likely to be assigned to the same group at a given BS. For instance, users 1, 2, and 3 are clustered within a single group under BS 1. Furthermore, Fig. \ref{Visualization}(b) illustrates the dynamic reconfiguration of BS-user connections in response to user mobility. In particular, when users 2 and 3 move, they are connected to BS 3. This is because each user is connected to the BS that is relatively closer to effectively utilize the advantages of proximity.
Fig. \ref{Visualization}(c) further demonstrates that geographically proximate users requesting the same content are not strictly assigned to the same BS. For instance, while users 4, 5, and 8 are served by BS 2, user 9 is associated with BS 3. This discrepancy arises because the signal reception directions for these users diverge significantly. If they are assigned to the same BS, the BS needs to transmit a wider beam to serve all users. However, the beamforming gain of the wider beam is lower, which reduces the sum rate of whole system.

\begin{figure}[!t]
	\centering
	\includegraphics[width=9cm]{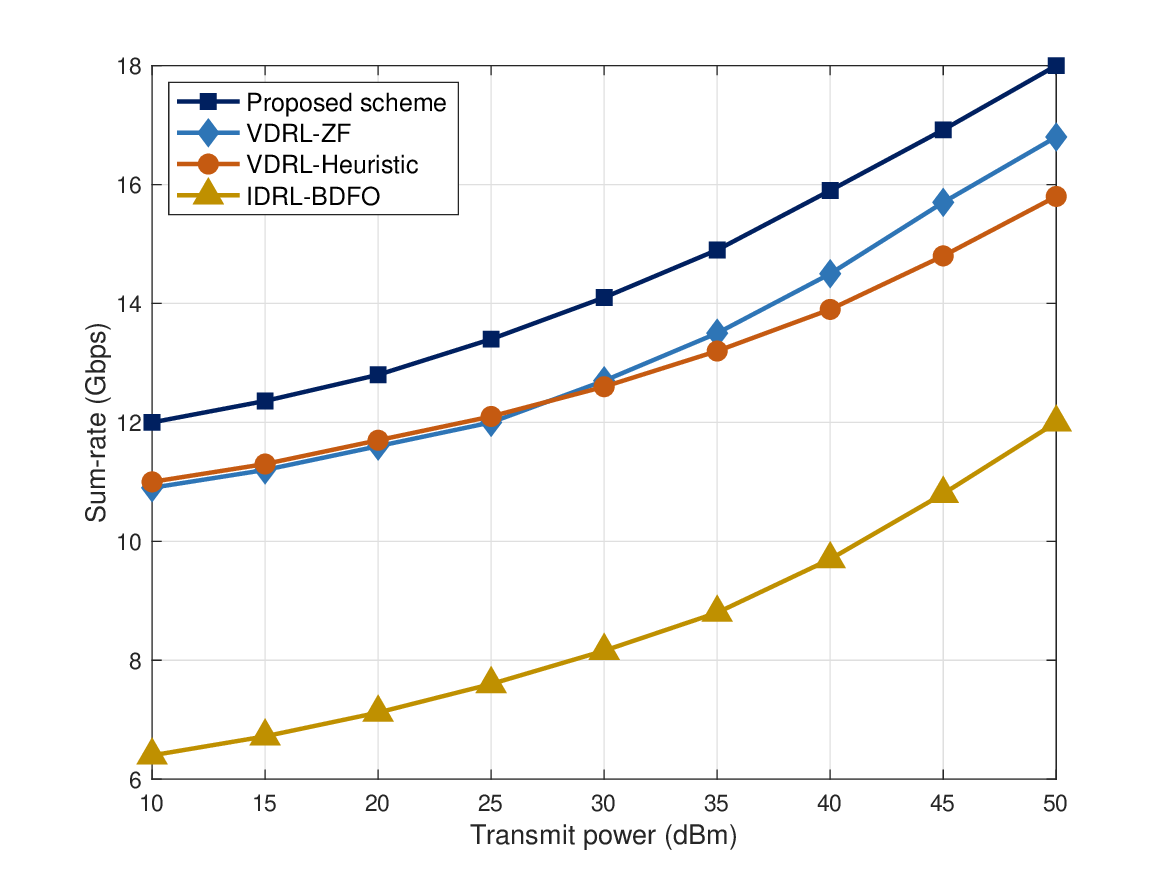}\\
	\caption{Sum-rate changes as the transmit power of each BS.} \label{RateTP}
\end{figure}

Fig. \ref{RateTP} illustrates the sum rate of all  multicast groups versus the transmit power of each BS. As shown, at a transmit power of 50 dBm, the proposed scheme achieves sum rate gains of up to 7.14\%, 13.9\%, and 50\% over VDRL-ZF, VDRL-Heuristic, and IDRL-BDFO, respectively. The 7.14\% and 13.9\% improvements over VDRL-ZF and VDRL-Heuristic are attributed to the proposed scheme's ability to dynamically balance interference suppression with desired signal enhancement. It is worth noting that VDRL-ZF exhibits inferior performance in the low-power regime due to its severe signal attenuation caused by strict zero-forcing constraints. However, as the transmit power increases, the network transitions from a noise-limited regime to an interference-limited regime. In this high-power regime, VDRL-ZF outperforms VDRL-Heuristic by completely eliminating inter-group interference, whereas the sum-rate growth of VDRL-Heuristic is significantly bottlenecked by severe interference leakage. The 50\% gain stems from the fact that the agents in the proposed scheme can collaborate to accomplish tasks in order to maximize the common goal, while, in the IDRL-BDFO scheme, the BSs greedily maximize their individual goals.

\begin{figure}[!t]
	\centering
	\includegraphics[width=9cm]{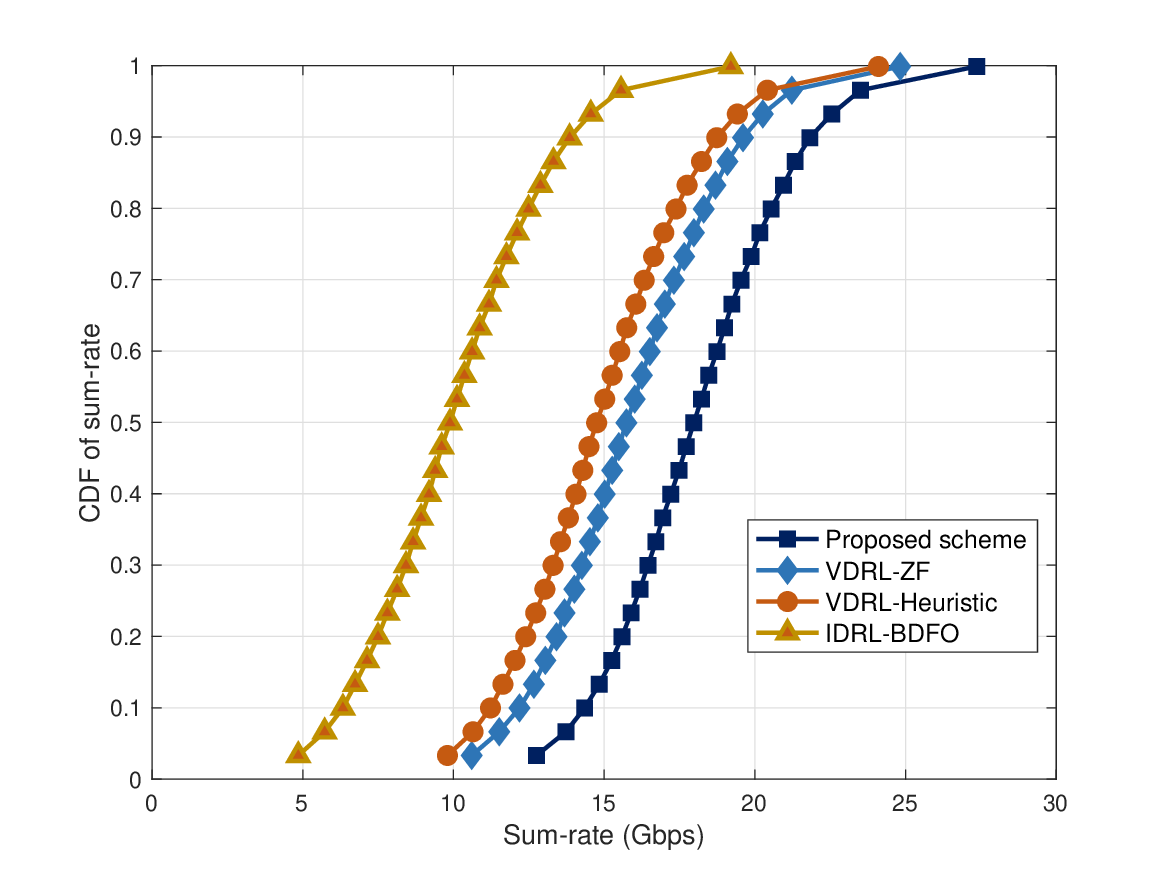}\\
	\caption{CDF of sum-rate.} \label{CDF}
\end{figure}

Fig. \ref{CDF} illustrates the cumulative distribution function (CDF) of the  multicast groups' sum-rate for a fixed BS transmit power of 50 dBm.  As observed, at a sum-rate of 15 Gbps, the proposed scheme achieves performance gains of up to 35.8\% and 56.5\% over VDRL-ZF and VDRL-Heuristic, respectively. This performance gap is attributed to the proposed algorithm's ability to balance desired signal enhancement with interference suppression. Specifically, it effectively mitigates both inter-group and intra-user inter-stream interference without significantly compromising the spatial degrees of freedom.  Meanwhile, for multicast scenarios, we design corresponding beamforming matrices for each group of users to maximize the total gain. In contrast, although VDRL-ZF slightly outperforms VDRL-Heuristic by strictly nulling the interference, it suffers from severe desired signal attenuation due to the exhaustive depletion of spatial degrees of freedom. 
Furthermore, Fig. \ref{CDF} shows that the proposed scheme achieves a CDF gain of up to 80.5\% over IDRL-BDFO at a sum-rate of 15 Gbps.
This huge advantage arises because the proposed scheme, empowered by value decomposition, can successfully find an ideal team policy to maximize the global team reward and handle the non-stationarity of the environment. The IDRL-BDFO algorithm, however, only finds a greedy policy that maximizes each BS's individual reward while ignoring the severe mutual interference among different BSs.

\begin{figure}[!t]
	\centering
	\includegraphics[width=9cm]{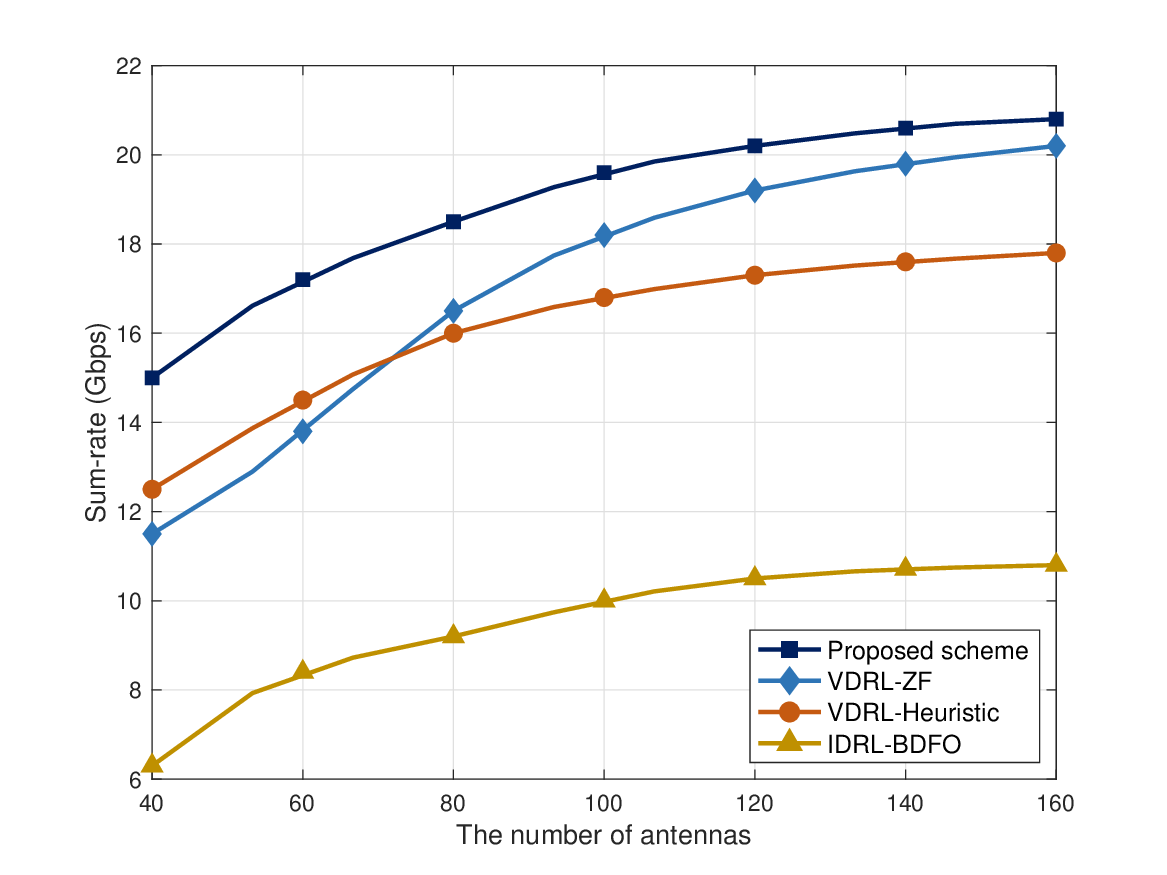}\\
	\caption{Sum-rate versus  the number of antennas.} \label{antennas}
\end{figure}

Fig. \ref{antennas} illustrates the sum rate of  multicast groups versus the number of antennas at each BS. As observed, the sum rate for all  algorithms increases monotonically with the antenna number. This upward trend is attributed to the fact that deploying more antennas at the BS yields higher array gains and provides more abundant spatial degrees of freedom.  Consequently, this significantly improves the interference suppressing capability and enhances the overall rate performance. 
It is particularly interesting to observe the crossover behavior between VDRL-ZF and VDRL-Heuristic. When the number of antennas is relatively small, the spatial degrees of freedom are  restricted. VDRL-ZF exhausts almost all available degrees of freedom to forcefully eliminate interference, resulting in severe desired signal attenuation and rendering its performance even inferior to VDRL-Heuristic. However, as the system scales up to massive MIMO configurations, the abundant spatial degrees of freedom allow VDRL-ZF to completely nullify the interference while still preserving sufficient degrees of freedom to enhance the desired signal. Thus, its sum rate surges and surpasses that of VDRL-Heuristic, which is persistently bottlenecked by unmanaged interference leakage. 
Furthermore, when the number of antennas of each BS reaches 160, the proposed algorithm achieves sum rate gains of up to 2.97\%, 16.85\%, and 92.59\% over VDRL-ZF, VDRL-Heuristic, and IDRL-BDFO, respectively. These improvements are attributed to the proposed scheme's ability to dynamically exploit spatial degrees of freedom, effectively balancing desired signal enhancement with interference suppression. Concurrently, it leverages value decomposition to discover an ideal team policy that maximizes the global reward.

\begin{figure}[!t]
	\centering
	\includegraphics[width=9cm]{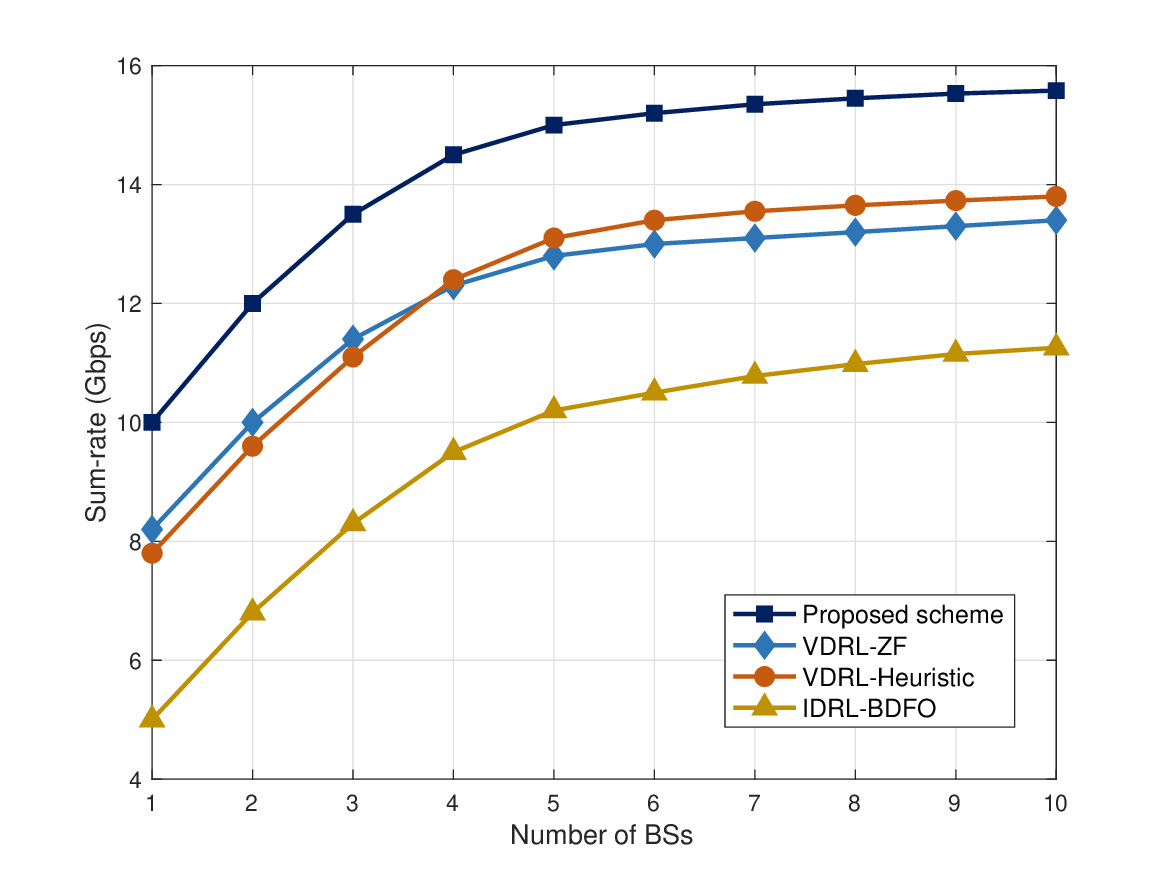}\\
	\caption{Sum-rate versus  the number of BSs.} \label{numberofBSs}
\end{figure}

Fig. \ref{numberofBSs} illustrates the sum rate of the  multicast groups versus the number of deployed BSs. As observed, the sum rate monotonically increases with a denser BS deployment. This improvement is primarily due to the fact that
 more BSs offer users a greater range of options, including the choice of which BS to connect to, as well as various beamwidths and beam directions. Therefore, a greater number of BSs can provide users with better coverage and improve the SINR of the system. Fig. \ref{numberofBSs} also illustrates that as the number of deployed BSs grows, the sum rate of  multicast groups rises at a slower pace. This is because there are already enough BSs to serve the users, and the increase in the number of BSs no longer significantly improves the channel gain and the beamforming gain. More importantly, the denser deployment of BSs severely exacerbates the inter-BS interference. As observed, VDRL-ZF slightly outperforms VDRL-Heuristic when the number of BSs is small. However, as the network becomes denser, VDRL-ZF suffers a severe performance bottleneck and is surpassed by VDRL-Heuristic. This is because the escalating inter-BS interference forces VDRL-ZF to severely deplete its spatial degrees of freedom for interference nulling, leading to significant desired signal attenuation. 
Furthermore, Fig. \ref{numberofBSs} reveals that with a deployment of 6 BSs, the proposed scheme achieves significant sum-rate improvements of up to 16.92\%, 13.43\%, and 44.76\% over VDRL-ZF, VDRL-Heuristic, and IDRL-BDFO, respectively. These substantial gains are fundamentally driven by the cooperative learning framework. Specifically, by exchanging individual $Q$-values, the BSs successfully acquire a global $Q$-value. Then, each BS can update its own parameters based on the global $Q$ value, which allows all BSs to collaborate in determining the users to be served. Moreover, the proposed algorithm strikes a balance by eliminating inter-group interference and intra-stream interference without severely sacrificing signal power, hence significantly improving the sum rate of all  multicast groups.

\begin{figure}[!t]
	\centering
	\includegraphics[width=9cm]{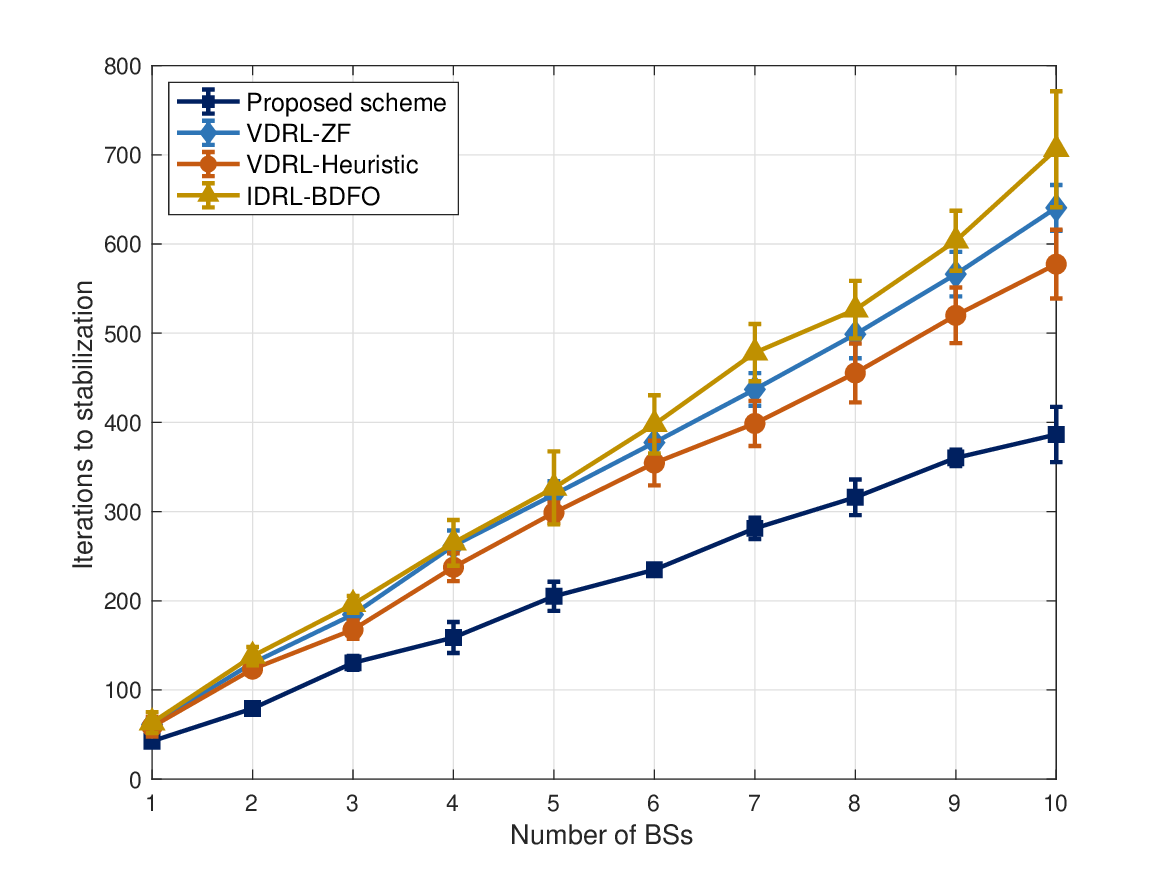}\\
	\caption{Convergence of the considered algorithms as the number of BSs varies. }\label{Convergence}
\end{figure}

Fig.~\ref{Convergence} illustrates how the mean number of iterations
required for empirical stabilization of the considered algorithms
changes as the number of BSs varies. The error bars represent
95\% confidence intervals across independent training runs.
As observed, increasing the number of BSs leads to a higher
average number of iterations required for empirical stabilization.
This is primarily because a larger network increases the complexity
of learning the BS-user association policies. Furthermore,
Fig.~\ref{Convergence} shows that the average number of iterations
required for empirical stabilization exhibits an approximately
linear increasing trend over the considered range of BS numbers.
In our proposed algorithm, each BS determines the users to be served
based on its local observations, while value decomposition enables
cooperative decision-making among BSs, thereby facilitating
the learning of association policies.
Furthermore, we can observe that VDRL-ZF requires more iterations
to achieve empirical stabilization than VDRL-Heuristic.
This can be attributed to the strict zero-forcing constraints
in VDRL-ZF, which can significantly reduce the available spatial
degrees of freedom and affect the resulting rewards, thereby
making the learning of association policies more challenging.
As further observed in Fig.~\ref{Convergence}, the proposed
algorithm requires significantly fewer iterations to achieve
empirical stabilization compared with the other three schemes.
This improvement can be attributed to the cooperative
decision-making mechanism of the proposed VD-RL algorithm,
which enables the BSs to coordinate their association decisions
through value decomposition. In contrast, IDRL-BDFO independently
learns the association policies of individual BSs, which 
result in slower empirical stabilization due to the
non-stationarity introduced by the interactions among BSs.

\section{Conclusions}

In this paper, we have investigated the sum-rate maximization problem within a multi-BS cooperative mmWave multicast system. Under this framework, the continuous mobility of users necessitates dynamic cooperation among multiple BSs to accommodate the time-varying service demands of mobile users. The problem has been cast into a unified optimization framework, and the VD-RL algorithm has been developed as an effective solution. Simulation results extensively validate the superiority of the proposed VD-RL scheme, which consistently outperforms conventional multi-agent schemes in terms of both convergence and achievable sum-rate.

\bibliographystyle{IEEEtran}
\bibliography{IEEEabrv,MMM}

\end{document}